\documentclass[manuscript,screen]{acmart}

\AtBeginDocument{%
  }
    
\begin{abstract}
Recently, a number of simulation testing approaches have been proposed to generate diverse driving scenarios for autonomous driving systems (ADSs) testing. 
\revision{However, many existing search-based approaches primarily determine NPC behaviors before scenario execution, which limits their ability to model interactions that depend on traffic signals and the Ego vehicle's real-time behavior. As a result, some reported violations may be dominated by unreasonable NPC behaviors, reducing the effectiveness of finding violations induced by the ADS, while the vast search space of NPC behaviors also limits efficiency.} 
To address these limitations, 
\revision{we propose a novel search-based testing framework, \tool, to generate more violation scenarios induced by the ADS. Specifically, \tool enables NPC vehicles to make maneuver decisions and generate trajectories according to traffic signals and the real-time behavior of the Ego vehicle, using different driving strategies. \tool further integrates this dynamic behavior generation with a genetic algorithm-based scenario configuration generator to improve the search for Ego-induced violations.} 
We compare \tool with four state-of-the-art scenario-based testing approaches. Our evaluation has demonstrated that \tool increases the proportion of violations induced by the ADS among all reported violations, on average, by \todo{125.21\%}, and improves the number of discovered unique violation patterns induced by the ADS by at least \todo{39.71\%}. Besides, \tool reduces the time to find the first violation induced by the ADS and the average time to find one violation induced by the ADS by \todo{82.13\%} and \todo{65.70\%}, respectively. \revision{We further conduct ablation studies along with sensitivity analyses of key parameters, and demonstrate the robustness and portability of \tool.}

\end{abstract}

\ccsdesc[500]{Software and its engineering~Software testing and debugging}

\keywords{Autonomous Driving Systems, Simulation Testing, Scenario-Based Testing}

\setcopyright{acmlicensed}
\copyrightyear{2026}
\acmYear{2026}
\acmDOI{XXXXXXX.XXXXXXX}

\acmJournal{TOSEM}
\acmVolume{1}
\acmNumber{1}
\acmArticle{1}
\acmMonth{8}    

\usepackage{xspace}
\usepackage{tcolorbox}
\usepackage{enumitem}
\usepackage[ruled]{algorithm2e}
\usepackage{multirow}
\usepackage{graphicx}
\usepackage{tabularx}
\newcolumntype{C}[1]{>{\centering\arraybackslash}p{#1}}
\usepackage{colortbl}
\usepackage{graphicx}  
\usepackage{float}  
\usepackage{subfig}
\usepackage{makecell} 
\usepackage{color}
\usepackage{arydshln} 
\usepackage{caption}
\usepackage{wrapfig}
\usepackage{colortbl}
\usepackage{amsthm} 
\usepackage{amsmath}
\usepackage{adjustbox}
\usepackage{threeparttable}
\usepackage{stfloats}
\usepackage{pifont}
\usepackage{xurl}
\usepackage{amsfonts}

\usepackage{balance}

\theoremstyle{definition}
\newtheorem{definition}{Definition}

\newcommand{\ie}[0]{\textit{i.e.,}\xspace}
\newcommand{\eg}[0]{\textit{e.g.,}\xspace}

\newcommand{\todo}[1]{\textcolor{black}{#1}}

\newcommand{\revision}[1]{\textcolor{black}{#1}}

\makeatletter

\newcommand{\Rmnum}[1]{\expandafter\@slowromancap\romannumeral #1@}
\makeatother

\newcommand{\tool}{\textsc{DynNPC}\xspace}
\newcommand{\toolcon}{\textsc{DynNPC-Yield}\xspace}
\newcommand{\tooladv}{\textsc{DynNPC-Adver}\xspace}
\newcommand{\toolagg}{\textsc{DynNPC-Over}\xspace}

\begin{document}

\title{\tool: Finding More Violations Induced by ADS in Simulation Testing via Dynamic NPC Behavior Generation}


\author{You Lu}
\affiliation{%
\department{College of Computer Science and Artificial Intelligence}
  \institution{Fudan University}
  \city{Shanghai}
  \country{China}
}

\author{Yifan Tian}
\affiliation{%
\department{College of Computer Science and Artificial Intelligence}
  \institution{Fudan University}
  \city{Shanghai}
  \country{China}
}

\author{Kun Zhang}
\affiliation{%
\department{College of Computer Science and Artificial Intelligence}
  \institution{Fudan University}
  \city{Shanghai}
  \country{China}
}

\author{Dingji Wang}
\affiliation{%
\department{College of Computer Science and Artificial Intelligence}
  \institution{Fudan University}
  \city{Shanghai}
  \country{China}
}

\author{Bihuan Chen}
\affiliation{%
\department{College of Computer Science and Artificial Intelligence}
  \institution{Fudan University}
  \city{Shanghai}
  \country{China}
}

\author{Haowen Jiang}
\affiliation{%
\department{College of Computer Science and Artificial Intelligence}
  \institution{Fudan University}
  \city{Shanghai}
  \country{China}
}

\author{Qicai Chen}
\affiliation{%
\department{College of Computer Science and Artificial Intelligence}
  \institution{Fudan University}
  \city{Shanghai}
  \country{China}
}

\author{Kun Hu}
\affiliation{%
\department{College of Computer Science and Artificial Intelligence}
  \institution{Fudan University}
  \city{Shanghai}
  \country{China}
}

\author{Xin Peng}
\affiliation{%
\department{College of Computer Science and Artificial Intelligence}
  \institution{Fudan University}
  \city{Shanghai}
  \country{China}
}

\renewcommand{\shortauthors}{Lu et al.}



\maketitle

\section{Introduction}
In recent decades, there has been a significant escalation in both academic and industrial commitment towards the development of autonomous driving systems (ADSs). These systems hold considerable promise for improving road safety, alleviating traffic congestion, and enhancing overall transportation efficiency, thereby driving a transformative shift in the automotive industry~\cite{paden2016survey}. Despite the advancements made by leading companies such as Tesla, Waymo, and Uber, current ADSs still struggle with corner cases and exhibit erroneous behaviors due to the extremely complicated real-world driving environments. These flaws in ADSs can lead to serious consequences and substantial losses, as highlighted by numerous documented traffic incidents~\cite{tesla, waymo, uber}. Consequently, extensive testing is needed to ensure the safety and reliability of ADSs before their deployment in the real world. 

Leading companies have employed on-road testing to evaluate the performance of ADSs. However, autonomous vehicles have to be driven more than 11 billion miles to demonstrate with 95\% confidence that autonomous vehicles are 20\% safer than human drivers~\cite{kalra2016driving}. This is not only time-consuming but also costly. In contrast, simulation testing offers a more efficient and cost-effective approach to generate diverse and challenging scenarios for ADSs by leveraging the high-fidelity simulators, such as LGSVL~\cite{lgsvl} and CARLA~\cite{Dosovitskiy17}. These simulators can generate a wide range of scenarios, including various weather conditions, road conditions, and traffic conditions. 


\revision{Several simulation testing approaches have been proposed for ADSs in simulators, and have been shown to be capable of finding violation scenarios. Some approaches propose domain-specific languages~\cite{Fremont2019, Queiroz2019, openscenario, opendrive, queiroz2024driver} to describe the scenarios, while others generate scenarios by reproducing real-world data~\cite{gambi2019generating,bashetty2020deepcrashtest,dai2024sctrans,zhang2023building}. In addition, many search-based approaches~\cite{Abdessalem2016a, Abdessalem2016b, li2020av, tian2022mosat, Tian2022, abdessalem2018testinga, Gladisch2020, luo2021targeting, kim2022drivefuzz, cheng2023behavexplor, huai2023doppelganger, sun2022lawbreaker, zhang2023testing, li2023simulation, Huai2023a, lu2023test, zhong2022neural} have been introduced to generate scenarios guided by different testing objectives. These search-based approaches typically predefine road conditions, weather conditions, and the behaviors of Non-Player Characters (NPCs) (\ie trajectories of maneuvers at each frame during simulation), and then iteratively mutate the resulting scenario configurations before executing them in the simulator. However, determining NPC behaviors before execution may insufficiently capture interactions conditioned on the Ego vehicle's real-time behavior and traffic signals during simulation, which can lead to unreasonable NPC behaviors. Moreover, exploring the vast search space of NPC behavior mutations is extensive and time-consuming, limiting the efficiency of finding violation scenarios. Recently, several works have explored the use of reinforcement learning to generate interactive and adversarial behaviors of NPCs~\cite{rowe2024ctrl, rempe2022generating, doreste2024adversarial, wang2025amacollision}. While these approaches enable reactive behaviors during execution, they usually rely on learned policies, which require substantial training data and computational resources. Besides, adversarial NPCs are often optimized to maximize the likelihood of violations, which may lead to behaviors that are overly aggressive or difficult to control and interpret.}

\revision{In fact, while an ADS should behave safely and appropriately even when encountering rule-breaking NPCs, it is unreasonable to expect it to handle completely unreasonable or physically impossible behaviors (\eg NPCs appearing in unreasonable locations, exhibiting abrupt and extreme speed changes, or violating traffic signals in ways that make collision avoidance impossible). As a result, violations found by scenario-based testing approaches may not necessarily reveal a bug in the ADS under test because the Ego vehicle (\ie the vehicle controlled by the ADS) may not bear the liability. This is also evidenced by a recent study~\cite{huai2023doppelganger}, where 1,109 violation scenarios are automatically generated in 240 hours. After manual diagnosis, all these violations are induced by NPC vehicles. Therefore, improving the effectiveness and efficiency of finding violation scenarios induced by the Ego vehicle, \ie \emph{Ego-induced violations}~(EIVs), remains a challenging problem in simulation testing for ADSs.}

\revision{In this work, we propose \tool, a novel search-based testing framework centered on runtime NPC behavior generation to find more EIVs effectively and efficiently. Specifically, \tool enables NPC vehicles to make maneuver decisions and generate trajectories online during simulation execution according to traffic signals and the real-time behavior of the Ego vehicle, under different driving strategies (\ie the yielding strategy, the adversarial strategy and the overtaking strategy). To support systematic testing, \tool further integrates a genetic algorithm-based generator to produce scenario configurations except for NPC trajectories, and a dedicated scenario executor to ensure the correct execution of generated scenarios.}

We have conducted large-scale experiments to evaluate the effectiveness and efficiency of \tool. We implement \tool based on Apollo 8.0 ~\cite{apollo} and LGSVL 2021.3~\cite{lgsvl} and compare it with four state-of-the-art scenario-based testing approaches (\ie \textsc{AV-Fuzzer}~\cite{li2020av}, \textsc{AutoFuzz}~\cite{zhong2022neural}, CRISCO~\cite{Tian2022}, and BehAVExplor~\cite{cheng2023behavexplor}). 
\revision{Our experiments demonstrate that \tool generates the most EIVs in 12 hours among the compared approaches in the straight road and crossroad scenarios, increases the proportion of EIVs among all reported violations, on average, by 125.21\%, and improves the number of discovered unique EIV patterns by at least 39.71\%. \tool also reduces the time to find the first EIV and the average time to find one EIV by 82.13\% and 65.70\%, respectively. Besides, \tool generates smoother and more diverse speed sequences for NPC vehicles. In addition, we analyze the effects of the genetic algorithm-based generator and the three NPC driving strategies on the testing results, study the sensitivity of key maneuver-constraint parameters. Finally, we validate the portability of \tool in a more complex roundabout scenario using CARLA.}

In summary, this work makes the following main contributions.
\begin{itemize}[leftmargin=*]
    \item We dynamically generate the behaviors of NPC vehicles using different driving strategies during simulation execution based on traffic signals and the real-time behavior of the Ego vehicle.
    \item \revision{We design and implement \tool, a novel search-based testing framework that integrates runtime NPC maneuver decision, online trajectory generation, scenario generation, and scenario execution to improve the possibility of finding Ego-induced violations in ADS simulation testing.}
    \item We conduct experiments with four state-of-the-art scenario-based testing approaches to demonstrate \tool's effectiveness and efficiency in finding violation scenarios induced by ADSs.
\end{itemize}

\section{Preliminary and Motivation} \label{sec:preliminary}
According to~\cite{ulbrich2015defining}, a scenario in ADS simulation testing refers to a collection of actors including the Ego vehicle attached with driving tasks, other traffic participants, \ie Non-Player Characters (NPCs) with behaviors over a period of time, and the environment (\eg weather conditions and traffic signal configurations). A scenario can be defined as follows. 

\begin{definition}
    A scenario $S =  \langle t^{S}, W,  E, \mathbb{T}, \mathbb{N} \rangle $ is a 5-tuple where:
    \begin{itemize}[leftmargin=*]
        \item $t^{S}$ is the maximum allowed frame duration for the scenario. 
        \item $W = \langle rain, fog, wetness, cloudness, time \rangle$ is a tuple used to specify \revision{weather} and the time of the day. $rain, fog, wetness$ and $cloudness$ are float numbers ranging from 0 to 1, and $time$ is an integer between 0 and 23.
        \item $E =  \langle p_{start}, p_{des} \rangle$ is a tuple indicating the driving task of Ego vehicle controlled by ADS under test, consisting of the starting position $p_{start}$ and the destination $p_{des}$.
        \item $\mathbb{T} = \{ T_{0}, T_{1}, \dots, {T_{\mid \mathbb{T} \mid -1}} \}$ is a set of traffic signal configurations with cooperative and mutually exclusive relationships, mapping the traffic signal $T_{k}$ to its color $c \in \{ RED, YELLOW, GREEN \}$ on a given map.
        \item $\mathbb{N} = \{ N_{0}, N_{1}, \dots, {N_{\mid \mathbb{N} \mid -1}}\}$ is a set of NPCs with behaviors that can be indicated as trajectories in a simulation. 
    \end{itemize}
\end{definition}


\begin{definition}
    \revision{The traffic signal configuration of $T_j = \langle c^{init}_{T_j}, c^{final}_{T_j},$ $ d^{init}_{T_j}, d^{trans}_{T_j}, d^{buffer}_{T_j} \rangle $ is a 5-tuple used to describe the temporal evolution of a traffic signal from its initial color to its final color where:}
    \begin{itemize}[leftmargin=*]
        \item \revision{$c^{init}_{T_j}$ is the starting signal color, $ c^{final}_{T_j}$ is the final signal color after the configured evolution, and $c^{init}_{T_j}, c^{final}_{T_j} \in \{RED, YELLOW, GREEN\}$. $d^{init}_{T_j}$ is the duration that the signal displays the color $c^{init}_{T_j}$.}
        \item \revision{$d^{trans}_{T_j}$ is the duration of the transition phase, if needed, when the signal evolves from $GREEN$ to $RED$, it displays $YELLOW$ for $d^{trans}_{T_j}$ to warn vehicles before turning $RED$~\cite{trafficsignal}.}
        \item \revision{$d^{buffer}_{T_j}$ is the safety buffer duration, if needed, during which the signal remains $RED$ before switching to $GREEN$, allowing vehicles that have already entered the intersection to clear it~\cite{trafficsignal}.}
    \end{itemize}
\end{definition}

\begin{definition}
    The trajectory of the Ego vehicle $E$ or an NPC $N_k \in \mathbb{N}$~is~a~2-tuple $\langle P, V  \rangle$, where:
    \begin{itemize}[leftmargin=*]
        \item $P =  \langle p^{0}, p^{1}, \dots, p^{d-1}  \rangle$ is a sequence of waypoints that the traffic participant follows~at each timestamp during the frame duration $d$. A waypoint $p$ indicates a specific location on the map in the coordinate system. 
        \item $V =  \langle v^{0},v^{1}, \dots, v^{d-1}  \rangle$ is a sequence of speed of the traffic participant at each timestamp during the frame duration $d$.
    \end{itemize}
\end{definition}

Scenario-based testing~\cite{zhong2021survey} configures the scenarios and executes them in a simulator bridged with the ADS under test. Then, the search problem for violation scenarios can be cast to a (multi-objective) optimization problem guided by the objective function that  formalizes a specific search goal, such as the minimal distance between vehicles and minimal \emph{time-to-collision}~\cite{tian2022mosat}. Previous approaches mutate the scenario configurations iteratively to change the weather conditions $W$, driving tasks of the Ego vehicle $E$, trajectories of NPCs $\mathbb{N}$ ~\cite{li2020av,kim2022drivefuzz}, and traffic signal configurations $\mathbb{T}$~\cite{huai2023doppelganger} in a vast searching space. Besides, some works~\cite{tian2022mosat, Tian2022} compose multiple trajectories of NPCs as behavior patterns to form the scenario configuration and choose to mutate these patterns, compressing the search space to improve efficiency. All of these approaches share the common feature of generating NPC behaviors in the scenario configuration prior to simulation execution. These approaches have been demonstrated to be capable of finding violation scenarios. However, the ADS may not bear the liability for the violations because the violations may be induced~by the NPCs. 

\begin{figure}
    \centering
    \subfloat[Example 1]{
        \includegraphics[width=0.1\linewidth]{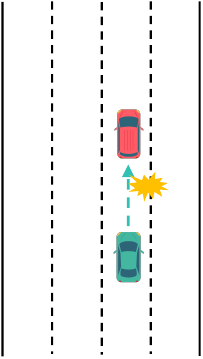}
        \label{img:NPC1}
    }
    \hspace{2cm}
    \subfloat[Example 2]{
        \includegraphics[width=0.1\linewidth]{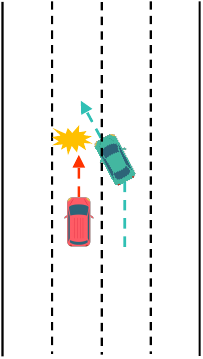}
        \label{img:NPC2}
    }
    \hspace{2cm}
    \subfloat[Example 3]{
        \includegraphics[width=0.185\linewidth]{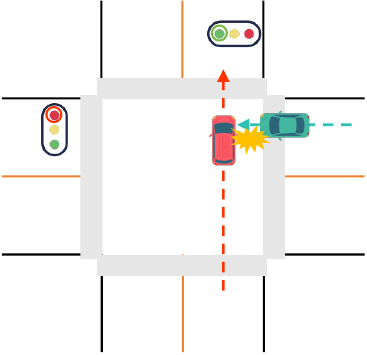}
        \label{img:NPC3}
    }
    \caption{Examples of Violations Induced by NPCs}
    \label{NPC Fault}
\end{figure}

We introduce the following motivating examples to demonstrate the limitation. We use the red car to represent the Ego vehicle and green car to represent the  NPC vehicle.  As shown in Fig.~\ref{img:NPC1}, the NPC vehicle hits the stopped Ego vehicle from behind. According to the Uniform Vehicle Code (UVC)~\cite{national1952uniform}, the rear vehicle is generally responsible in rear-end collisions. In Fig.~\ref{img:NPC2}, the NPC vehicle suddenly changes lanes with an abrupt speed change over a short distance in front of the Ego vehicle and the Ego vehicle decelerates but still fails to avoid the collision. In Fig.~\ref{img:NPC3}, the Ego vehicle moves forward because it recognizes that the traffic signal of its current lane is green and the intersection is clear. When the Ego vehicle passes the intersection, it is collided by a high-speed NPC vehicle that does not obey the red signal in the horizontal direction. In fact, all the aforementioned violations reported by previous approaches are induced by the unreasonable NPC behaviors. 
These approaches ignore the behavior of the Ego vehicle during scenario execution and inevitably introduce unreasonable NPC behaviors generated through iterative mutations prior to execution, \revision{resulting in} a high false positive rate of reported violations in the testing results.

To address the above limitation while facilitating the search for EIVs, we propose \tool to dynamically generate the behaviors of NPC vehicles based on traffic signals and the real-time behavior of the Ego vehicle during simulation execution, finding more EIVs effectively and efficiently.


\begin{figure}[!t]
    \centering
    \includegraphics[width=0.7\linewidth]{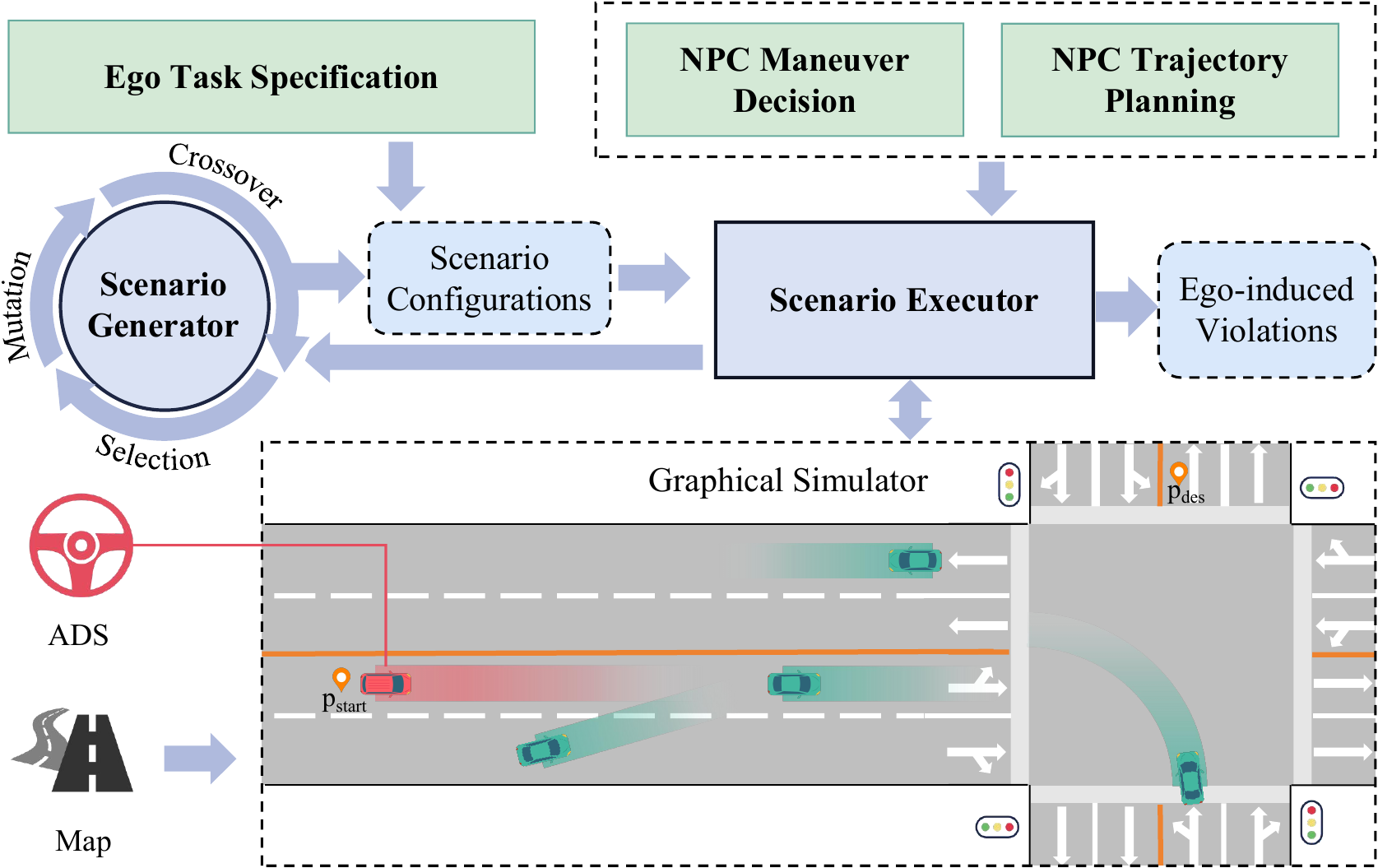}
    \caption{Approach Overview of \tool}
    \label{img:overview}
\end{figure}

\section{Methodology}

During the simulation execution, NPC vehicles can have a variety of behaviors, and different driving maneuvers may have different degrees of impact on the Ego vehicle. Besides, NPC vehicles should obey traffic signals when taking maneuvers at intersections equipped with traffic lights. 
\revision{Even under the same maneuver, the generated trajectories may still differ in their speed profiles. For example, when an NPC vehicle changes lanes in front of the Ego vehicle, it may either slow down to yield before completing the lane change or speed up to overtake in front of the Ego vehicle. It may adopt an adversarial driving strategy, posing a greater threat to the Ego vehicle.
Therefore, the key challenge~for dynamically generating the behaviors of NPC vehicles during simulation execution lies in \emph{(1) how to determine the maneuvers of the NPC vehicles and (2) how to generate the trajectories obeying traffic signals using different~driving~strategies}.}

\revision{Fig.~\ref{img:overview} shows the overview of \tool. First, \tool specifies the driving task of the Ego vehicle (see Sec.~\ref{sec:ego_task}). Then, it enables NPC vehicles to make maneuver decisions based on the real-time behavior of the Ego vehicle (see Sec.~\ref{sec:maneuver_decision}) and to generate trajectories under different driving strategies while obeying traffic signals (see Sec.~\ref{sec:trajectory_generation}). To improve the effectiveness and efficiency of finding EIVs, \tool connects the ADS with the simulator and selects a road segment from the map loaded in the simulator. It further employs a genetic algorithm-based scenario generator (see Sec.~\ref{sec:scenario_generator}) to automatically search scenario-level configurations in our genetic representation. Finally, \tool implements a new scenario executor (see Sec.~\ref{sec:simulation_executor}) to ensure correct scenario execution and collect testing results.}

\subsection{Ego Task Specification}\label{sec:ego_task}
As mentioned in Sec.~\ref{sec:preliminary}, the driving task of the Ego vehicle can be represented by the starting position $p_{start}$ and the destination $p_{des}$. The goal of the Ego vehicle is to navigate from  $p_{start}$  to  $p_{des}$  while considering dynamic traffic conditions and adhering to traffic signals. We select specific road segments from the provided map, supporting both straight roads and crossroads controlled by traffic lights. A set of $p_{start}$ and $p_{des}$ extracted from the map are assigned to ensure a diverse range of driving tasks.
For straight road scenarios, multiple NPC vehicles are initially distributed within the road segment between  $p_{start}$  and $p_{des}$. These NPC vehicles may occupy different lanes and follow varying speeds, or perform lane changes during simulation execution, creating a dynamic environment for the Ego vehicle. The Ego's task is to pass through this region, ensuring safe interactions with NPC vehicles while reaching $p_{des}$. 

For crossroad scenarios, NPC vehicles are initially positioned in designated waiting areas before the stop lines at the crossroad. The Ego vehicle is required to navigate the crossroad by either going straight, turning left, or turning right, based on the assigned  $p_{start}$ and $p_{des}$. During this process, the Ego vehicle must strictly comply with traffic signals and avoid potential conflicts with crossing NPC vehicles.

\subsection{NPC Maneuver Decision}\label{sec:maneuver_decision}
According to the data provided by National Highway Traffic Safety Administration (NHTSA)~\cite{NHTSA}, in real-word driving scenarios, vehicles can perform various maneuvers including decelerating, accelerating, lane changing, turning left, turning right, parking, and backing up,~\cite{najm2013description, botzer2019relationship}. We introduce the constraints of maneuvers we adopt and illustrate how NPC vehicles in \tool dynamically determine their driving maneuvers based on the behavior of the Ego vehicle.

\textbf{Maneuver Constraints.} \revision{\tool adopts a maneuver taxonomy based on CRISCO~\cite{Tian2022}, but our maneuver constraints are not directly reused from CRISCO. CRISCO mainly constrains the geometric and topological feasibility of predefined behavior patterns, such as the relative positions of start and destination points, lane relationships, and movement directions. In contrast, \tool introduces additional runtime-oriented constraints for maneuver feasibility, including the longitudinal safety distance to the Ego vehicle, the activation of brake/turn signals, solid-line checking for lane changes, lane-type constraints for turning maneuvers, traffic-signal compliance, and speed-related constraints consistent with the ADS under test. These constraints are introduced to ensure that NPC vehicles do not appear in unreasonable positions, perform abrupt lane changes within short distances that could lead to unavoidable collisions, and comply with traffic signals so that their behaviors remain as reasonable as possible. By enforcing such constraints during runtime maneuver decision, we aim to reduce unreasonably designed scenarios where collisions are primarily caused by improper NPC behaviors rather than by the Ego vehicle.}

\textit{Decelerating.}
When the NPC vehicle $N_k$ decides to decelerate at frame $t_0$, it must ensure a safe deceleration longitudinal distance to the Ego vehicle behind if they are in the same lane, and activate the brake lights, avoiding sudden braking and rear-end collision. The specification can be defined by Eq.~\ref{eq:decelerate},
\begin{equation}\label{eq:decelerate}
    \begin{array}{c}
            D_{N2E}(p_{N_k}^{t_0}, p_{E}^{t_0}) \geq threshold 
            \wedge \ brakeSignal(p_{N_k}^{t_0}) = True
    \end{array}
\end{equation}
where $D_{N2E}(p_{N_k}^{t_0}, p_{E}^{t_0})$ returns the distance between the NPC vehicle and the Ego vehicle at frame $t_0$ when they drive in the same lane, \revision{and $threshold$ is set to 30 meters by default as a conservative safety threshold.} The function $brakeSignal(p_{N_k}^{t_0})$ returns the status of NPC vehicle's braking~signal. \revision{Besides, the maximum deceleration rate of the NPC vehicle is set consistently with the configuration of the ADS under test, \eg 8 $m/s^2$ of Apollo~\cite{apollo}.}

\textit{Accelerating.}
\revision{When the NPC vehicle $N_k$ decides to accelerate at frame $t_0$, if the NPC vehicle is behind the Ego vehicle in the same lane, it must maintain a safe longitudinal distance from the Ego vehicle. In this case, the maximum speed of the NPC vehicle after acceleration must not exceed the speed of the Ego vehicle, so as to avoid introducing an unreasonable rear-end collision. Besides, it should ensure that the speed after acceleration does not exceed the speed limit of the current road section. The specification can be defined by Eq.~\ref{eq:accelerate},}
\revision{\begin{equation}\label{eq:accelerate}
    \begin{array}{c}
        \big(0 < D_{N2E}(p_{E}^{t_0}, p_{N_k}^{t_0}) \le threshold \Rightarrow MaxSpeed_{N_k} \leq v_{E}^{t_0} \big) 
        \wedge \ MaxSpeed_{N_k} \leq \text{speedLimit}
    \end{array}
\end{equation}
}
where \revision{$D_{N2E}(p_{E}^{t_0}, p_{N_k}^{t_0})$} returns the distance between the Ego vehicle and the NPC vehicle at frame $t_0$ when they \revision{drive} in the same lane, \revision{$threshold$ is set to 30 meters by default as a conservative safety threshold,} and $\text{speedLimit}$ is extracted from the map. \revision{Besides, the maximum acceleration rate of the NPC vehicle is also set consistently with the configuration of the ADS under test, \eg 8 $m/s^2$ of Apollo~\cite{apollo}.}

\textit{Lane Changing.} 
When the NPC vehicle \(N_k\) is going to take lane changing maneuver to the left at frame \(t_0\), it must ensure a safe lane changing distance from the following Ego vehicle, activate~the left turn signal when changing lanes, and avoid performing the~maneuver within a solid-line area. The specification can be defined~by~Eq.~\ref{eq:left_change}, 
\begin{equation}\label{eq:left_change}
    \begin{array}{c}
            D_{N2E}(p_{N_k}^{t_0}, p_{E}^{t_0}) \geq threshold  
            \wedge \ leftSignal(p_{N_k}^{t_0}) = True \wedge \ isSolid(p_{N_k}^{t_0}) = False
    \end{array}
\end{equation}
where $D_{N2E}(p_{N_k}^{t_0}, p_{E}^{t_0})$ returns the longitudinal distance between the NPC vehicle and the Ego vehicle at frame $t_0$, and $threshold$ \revision{is set to 30 meters by default as a conservative safety threshold for lane-changing feasibility.} The function $leftSignal(p_{N_k}^{t_0})$ returns the status of NPC \revision{vehicle's} left turn signal, while $isSolid(p_{N_k}^{t_0})$ indicates whether the surrounding lane markings of the NPC vehicle at frame $t_0$ are solid lines.

Similarly, when NPC $N_k$ performs a right lane change at frame $t_0$, the specification is defined by Eq.~\ref{eq:right_change}.
\begin{equation}\label{eq:right_change}
    \begin{array}{c}
            D_{N2E}(p_{N_k}^{t_0}, p_{E}^{t_0}) \geq threshold  
            \wedge \ rightSignal(p_{N_k}^{t_0}) = True \wedge \ isSolid(p_{N_k}^{t_0}) = False
    \end{array}
\end{equation}
where \(rightSignal(p_{N_k}^{t_0})\) checks if the right turn signal is turned~on.

\textit{Turning Left.} When the NPC vehicle $N_k$ decides to turn left at an intersection at frame $t_0$, it must ensure that the maneuver is performed safely in the left turn lane, following traffic signals (see Speed Planning). Besides, it must activate the left turn signal before initiating the turn. The specification can be defined~by~Eq.~\ref{eq:left_turn},
\begin{equation}\label{eq:left_turn}
    \begin{array}{c}
        isLeftTurnLane(p_{N_k}^{t_0}) = True \wedge leftSignal(p_{N_k}^{t_0}) = True 
        \wedge RedSignalObeying = True
    \end{array}
\end{equation}
where $isLeftTurnLane(p_{N_k}^{t_0})$ ensures that the NPC vehicle only turns left in the left turn lane and $leftSignal(p_{N_k}^{t_0})$ ensures the activation of the left turn signal.

\textit{Turning Right.}
\revision{When the NPC vehicle $N_k$ decides to turn right at an intersection at frame $t_0$, it must ensure that the maneuver is performed safely in the right turn lane and activate the right turn signal before initiating the turn. In our setting, when no dedicated right-turn signal is present, a right turn is treated as a permissible maneuver even if the forward through signal is red according to traffic regulations~\cite{mutcd2003_signal_features, california_dmv_handbook_road_rules, signal}. The specification can be defined~by~Eq.~\ref{eq:right_turn},}
\begin{equation}\label{eq:right_turn}
    \begin{array}{c}
        isRightTurnLane(p_{N_k}^{t_0}) = True \wedge \ rightSignal(p_{N_k}^{t_0}) = True 
    \end{array}
\end{equation}
where \revision{$isRightTurnLane(p_{N_k}^{t_0})$} ensures that the NPC vehicle only turns right in the right turn lane and $rightSignal(p_{N_k}^{t_0})$ ensures the activation of the right turn signal.

\textit{Parking.} 
When the NPC vehicle chooses to decelerate to stop, it must ensure that the parking maneuver is performed in a designated parking area, at a complete stop at frame $t_n$. The specification can be defined by Eq.~\ref{eq:parking},
\begin{equation}\label{eq:parking}
    \begin{array}{c}
    prohibitParkingZone(p_{N_k}^{t_n}) = False 
    \wedge \ v_{N_k}^{t_n} = 0 
    \end{array}
\end{equation}
where $prohibitParkingZone(p_{N_k}^{t_n})$ returns whether the NPC vehicle is stopping in prohibited areas (\eg intersections).

\textit{Backing up.} We do not consider backing up maneuvers because it is the most dangerous maneuver in highways and is forbidden in the crossroad. 

\textbf{Maneuver Decision.}
For the NPC vehicle $N_k$, we determine its maneuver at frame $t$ based on the behavior of the Ego vehicle. Specifically, we obtain the speed \revision{$v_{E}^{t}$} of the Ego vehicle at frame $t$ from simulator, and its driving task (\ie $p_{start}$ and $p_{des}$). 
Then, we generate the expected trajectory of the Ego vehicle. 
On the straight road, the expected trajectory is generated according to the Ego vehicle's heading direction provided by the simulator, assuming uniform motion at the current speed \revision{$v_{E}^{t}$}, resulting in a straight trajectory for the next period of time.
On the crossroad, the expected trajectory is generated according to the driving task, forming a path that connects $p_{start}$ and $p_{des}$ while adhering to roadway centerline. 
Given the position $p_{N_{k}}^{t}$ of the NPC vehicle obtained from the simulator, we identify all feasible driving maneuvers that satisfy the maneuver constraints. 
Among these maneuvers, if there exists a sequence of waypoints generated by maneuver $m$ (see Sec.~\ref{sec:trajectory_generation}) that overlaps with the expected trajectory of the Ego vehicle, the NPC vehicle will execute maneuver $m$ in the subsequent time period to enhance its interaction with the Ego vehicle, thereby reducing the search space for EIVs. If none of the feasible maneuvers' waypoint sequences overlap with the Ego's expected trajectory, the NPC vehicle will randomly select one from the set of all available~maneuvers. 

Fig.~\ref{img:maneuver} shows an example of maneuver decision on the crossroad. According to the driving task, the Ego vehicle will proceed straight through the crossroad when its traffic signal is green. At this moment, the NPC vehicle A, positioned in the oncoming through and left-turn lane, can choose to either go straight or turn left. It will select the left-turn maneuver from the available options, as the trajectory of left-turn maneuver overlaps with the expected trajectory of the Ego vehicle, increasing the likelihood of interaction between the two vehicles. Meanwhile, the NPC vehicle B, located in the cross-direction through and right-turn lane, faces a red signal for through traffic due to the mutually exclusive signal configuration with the Ego vehicle's green signal. Under this configuration, NPC vehicle B is permitted to execute only a right turn maneuver.

\subsection{NPC Trajectory Planning}\label{sec:trajectory_generation}

As mentioned in Sec.~\ref{sec:preliminary}, a trajectory consists of a sequence of waypoints and its speed sequence. Thus, we divide trajectory planning process into two tasks, namely waypoint generation and speed generation. 

\textbf{Waypoint Generation.} The waypoint generation task is responsible for generating a sequence of waypoints that the NPC vehicle should follow to execute the selected maneuver. Similar to previous work~\cite{Tian2022, queiroz2024driver}, Bézier curves \cite{mortenson1999mathematics}, which are parametric curves providing a smooth transition, are used to calculate the waypoints for this phase. A Bézier curve $B(t)$ can be constructed by four control points $P_{0}-P_{3}$, i.e., $B(\zeta)=(1-\zeta)^{3}P_{0}+3(1-\zeta)^{2}\zeta P_{1}+3(1-\zeta)\zeta^{2}P_{2}+\zeta^{3}P_{3},\ \zeta \in [0,1]$. For example, when generating the waypoints of lane changing maneuver, we set the current position $p_{N_k}^{t_0}$ of the NPC vehicle as $P_0$, and the target lane position extracted from map as $P_3$. \revision{Then, we compute the two intermediate control points $P_1$ and $P_2$ according to the lane geometry. Specifically, let $\vec{h}_0$ and $\vec{h}_3$ denote the heading directions of the current lane at $P_0$ and the target lane at $P_3$, respectively. We place $P_1$ by moving forward from $P_0$ along $\vec{h}_0$, and place $P_2$ by moving backward from $P_3$ along $\vec{h}_3$, using the same longitudinal offset proportional (0.3$\times$) encoded in the waypoint specification~\cite{Tian2022} to the distance between $P_0$ and $P_3$. In this way, the Bézier curve follows the entering direction of the current lane and the exiting direction of the target lane, yielding a smooth and plausible lane-changing trajectory.}

\textbf{Speed Generation.} The speed generation task is responsible for assigning speeds to the generated waypoints. By default, NPC vehicles operate at their initially assigned speeds. However, strictly maintaining this speed may lead to running red lights on the crossroad. Additionally, when interacting with the surrounding Ego vehicle, adhering rigidly to the default speed may result in suicidal behaviors, potentially leading to aggressive collisions. 
\revision{Thus, we use the Station\text{-}Time graph (\ie $s \text{-} t$ graph)~\cite{Spatial-Temporal-Graph} to modify the speed profile of the NPC vehicle's generated waypoints. The $s\text{-}t$ graph has also been used in prior work such as ACAV~\cite{sun2024acav}, where it is adopted to analyze ADS planning states in accident recordings. In \tool, we use the $s\text{-}t$ graph for runtime speed planning of NPC vehicles. Specifically, in an $s\text{-}t$ graph, time is the horizontal axis, the planned longitudinal trajectory distance is the vertical axis, and the planned longitudinal trajectory is a curve. Each point on the curve represents a waypoint on the planned trajectory, and the curve's gradient represents the speed. Based on the generated waypoints, we use the $s\text{-}t$ graph to adjust the NPC vehicle's speed profile, so as to enable NPC vehicles to obey traffic signals and implement different driving strategies when their expected trajectories overlap with that of the Ego vehicle.}

\begin{figure}[!t]
    \centering
    \begin{minipage}[t]{0.48\textwidth}
        \centering
        
        \includegraphics[width=.9\textwidth, height = 3.5cm]{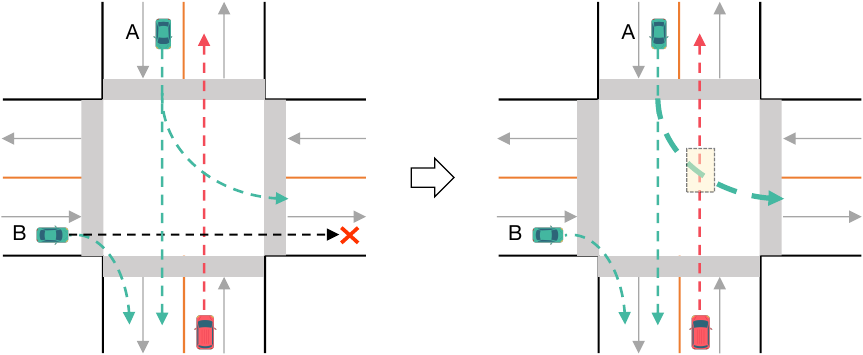}
        
        \caption{Examples of NPC Maneuver Decision on Crossroad}
        \label{img:maneuver}
    \end{minipage}%
    \hfill
    \begin{minipage}[t]{0.48\textwidth}
        \centering
        \includegraphics[width=.95\textwidth]{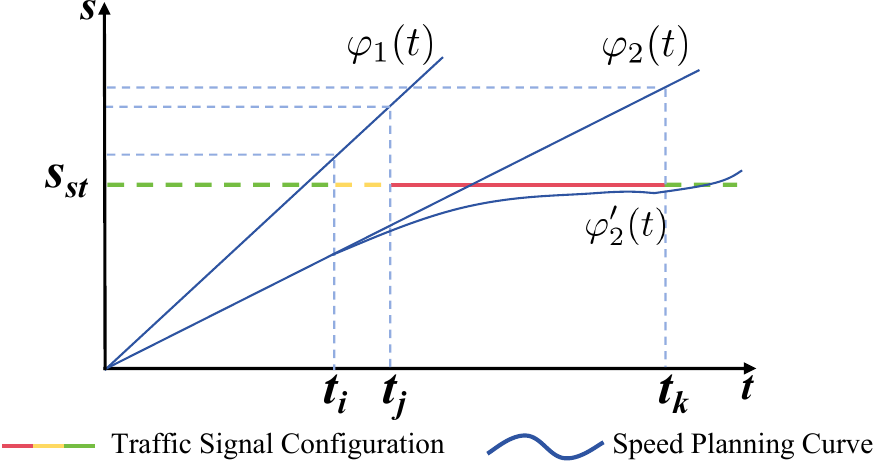}
        \caption{Traffic Signal Obeying with $s\text{-}t$ Graph}
        \label{img:signal}
    \end{minipage}
\end{figure}

\begin{figure}
    \centering
    \includegraphics[width=0.75\linewidth]{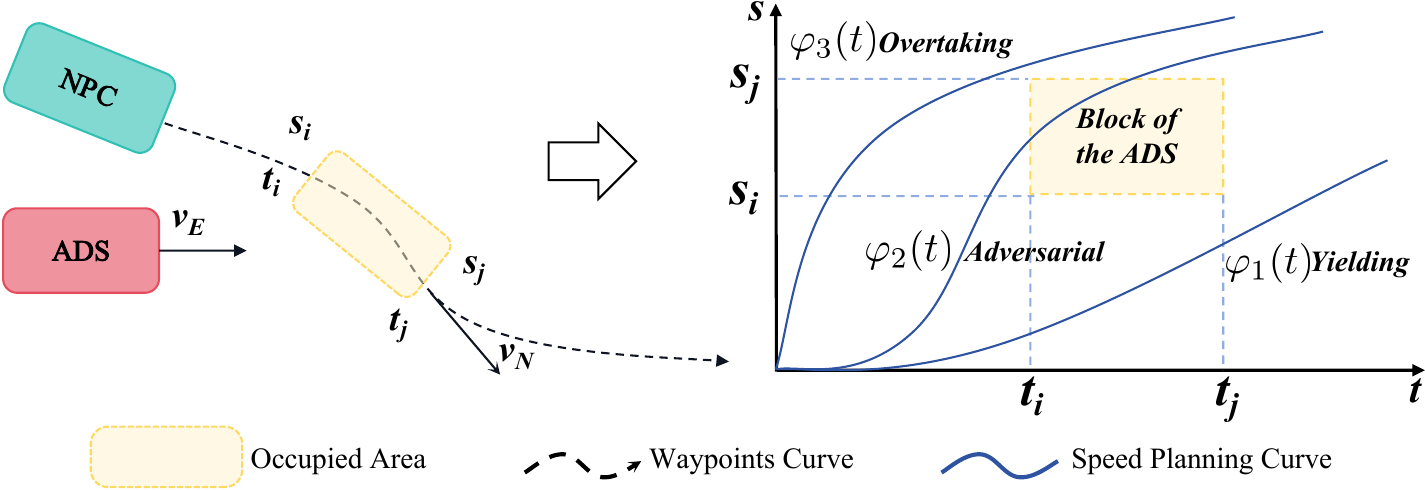}
    \caption{Driving Strategy Implementation with the $s\text{-}t$ Graph}
    \label{img:strategy}
\end{figure}

\emph{Traffic Signal Obeying.} When the NPC vehicle $N_k$ attempts to pass through the crossroad at frame $t$ with an initial speed $v_{N_k}^{t}$, we retrieve from the simulator both the NPC vehicle's current distance to the stop line, denoted as $s_{st}$, and the traffic signal configuration. As illustrated in Fig.~\ref{img:signal}, assuming that the traffic signal remains green from $t$ to $t_i$, turns yellow from $t_i$ to $t_j$, and becomes red from $t_j$ to $t_k$, we use $v_{N_k}^{t}$ as the slope to draw the speed planning curve of $N_k$. Here, $\varphi_1(t)$ means that $N_k$ can enter the crossroad before $t_i$ (\ie green signal) at the current speed. Conversely, $\varphi_2(t)$ means that the vehicle will pass the stop line after $t_j$, which will result in running a red light. To prevent this, we replan the speed profile of $N_k$, gradually decelerating until $N_k$ comes to a complete stop before the stop line (\ie the $\varphi'_2(t)$ curve). Finally, we update the speed of each waypoint based on the gradient of the newly generated speed planning curve, ensuring that the NPC vehicle adheres to traffic signals while executing its maneuver.

\emph{Driving Strategy Implementation.} When the waypoints of the NPC vehicle $N_k$ overlap with the expected trajectory of the Ego vehicle after frame $t$, we use the $s \text{-} t$  graph to generate \revision{the} speed of $N_k$, as shown in Fig.~\ref{img:strategy}. Assuming that the total length of the waypoint sequence generated by $N_k$'s maneuver $m$ is $s$, and the Ego vehicle will travel at a constant speed $v_{E}^{t}$ obtained from the simulator since frame $t$, then, the Ego vehicle will occupy the $s_i$ to $s_j$ section of $N_k$'s waypoints from time $t_{i}$ to $t_{j}$. Therefore, there is a block of the ADS in the $s\text{-}t$ graph where a collision is most likely to occur. Then, we can plan the speed of $N_k$ with three different driving strategies as follows.
\begin{itemize}[leftmargin=*]
    \item \textbf{Yielding Strategy:} The NPC vehicle decelerates and yields to the Ego vehicle, planning a speed profile below the block of the ADS (\eg $\varphi_1(t)$ in Fig.~\ref{img:strategy}).
    \item \textbf{Adversarial Strategy:} The NPC vehicle travels with adversarial speeds, planning a speed profile through the block of the ADS (\eg $\varphi_2(t)$ in Fig.~\ref{img:strategy}).
    \item \textbf{Overtaking Strategy:} The NPC vehicle accelerates to overtake the Ego vehicle, planning a speed profile above the block of the ADS (\eg $\varphi_3(t)$ in Fig.~\ref{img:strategy}).
\end{itemize}

Finally, we calculate the gradient of the speed planning curve and update the speed of each waypoint of $N_k$ to implement the selected driving strategy, improving the diversity of NPC behaviors.

\subsection{Scenario Generator}\label{sec:scenario_generator}
\revision{To systematically explore scenario-level configurations beyond runtime NPC trajectory generation, \tool adopts a genetic algorithm-based scenario generator similar in spirit to previous work~\cite{li2020av, Tian2022, tian2022mosat, abdessalem2018testinga, kim2022drivefuzz} to automatically generate scenario configurations for the ADS under test.} We elaborate on the design of the genetic representation of the scenario configurations and its search operators.

\begin{figure}[!t]
    \centering
    \includegraphics[width=0.6\linewidth]{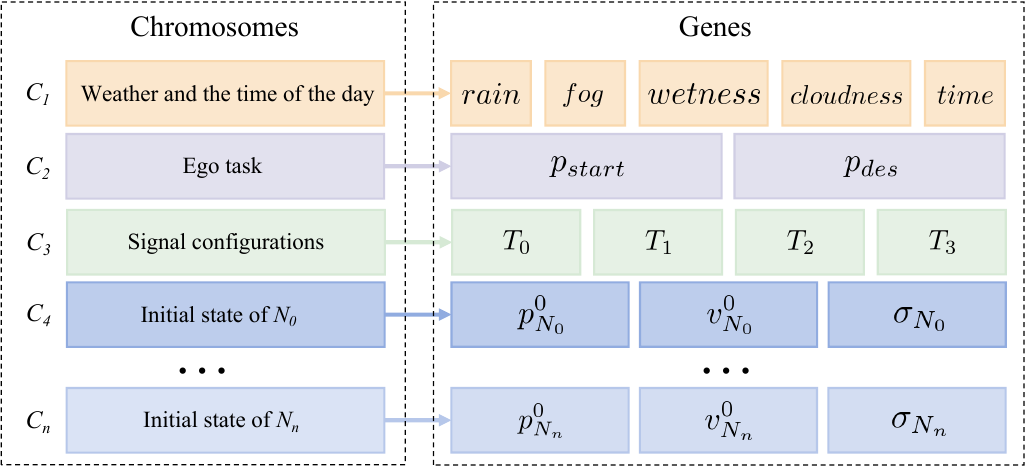}
    \caption{Genetic representation used by \tool}
    \label{img:genetic_representation}
\end{figure}

\textbf{Genetic Representation.} We use the definition of a scenario in Sec.~\ref{sec:preliminary} to configure scenarios. 
However, instead of explicitly specifying the trajectories of NPC vehicles in our scenario configurations $Conf$, we assign them driving strategies (\ie yielding strategy, adversarial strategy, or overtaking strategy).
Fig.~\ref{img:genetic_representation} illustrates a genetic representation of an individual generated by \tool. Each individual, which is a scenario configuration in our representation, consists of several chromosomes. $C_1$ indicates the weather conditions and the time of the day (\ie $W$), $C_2$ indicates the driving task of the Ego vehicle (\ie $E$), and $C_3$ indicates the traffic signal configurations at the crossroad (\ie $\mathbb{T}$). Besides, we use~chromosomes from $C_4$ to $C_n$ that have 3 genes (\ie the initial position $p^{0}_{N_{k}}$, the initial speed $v^{0}_{N_{k}}$ and the driving strategy $\sigma_{N_{k}}$ adopted by $N_k$) to indicate the NPC vehicles engaged in the scenario (\ie $\mathbb{N}$), respectively. 
\revision{This representation differs from prior genetic algorithm-based approaches in that it does not encode full NPC trajectories as genes. Instead, it only encodes scenario-level initial conditions and high-level driving strategies, while the detailed trajectories are generated online by the maneuver decision and trajectory planning modules during simulation execution.}

\begin{figure}
    \centering
    \subfloat[Uniform Mutation]{
        \includegraphics[width=0.47\linewidth]{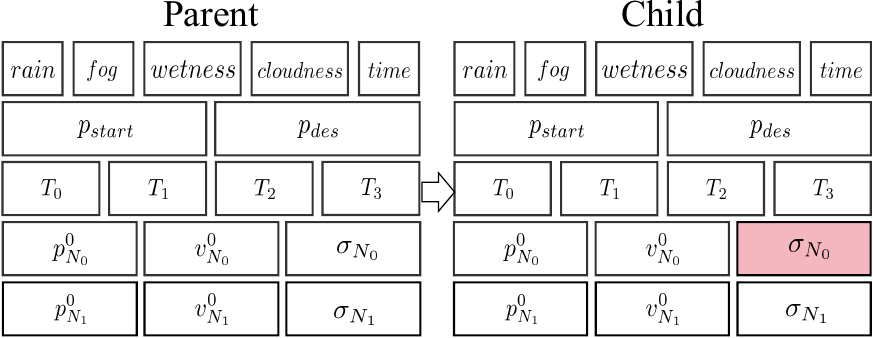}
        \label{img:Mutation1}
    }
    \subfloat[Gene Exchange]{
        \includegraphics[width=0.47\linewidth]{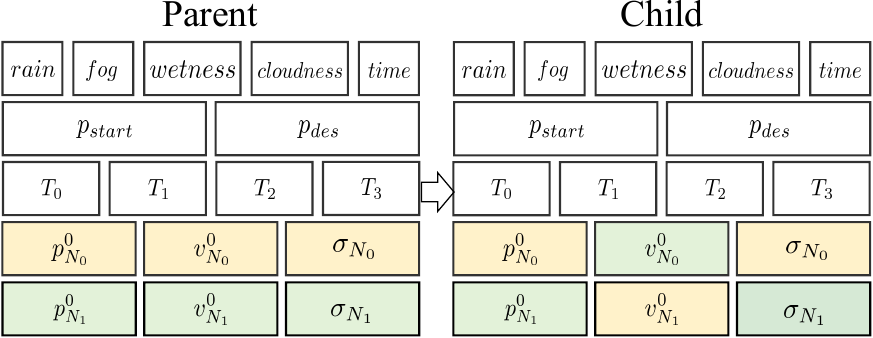}
        \label{img:Mutation2}
    }
    \caption{Examples of Mutation}
    \label{img:Mutation}
\end{figure}

\textbf{Mutation.} 
\revision{Mutation introduces intra-individual variation in our scenario representation. As shown in Fig.~\ref{img:Mutation}, we mainly apply uniform mutation to randomly change one gene in the chromosomes, and additionally exchange the same gene between two NPC chromosomes within one individual.}

\begin{figure}[!t]
    \centering
    \includegraphics[width=\linewidth]{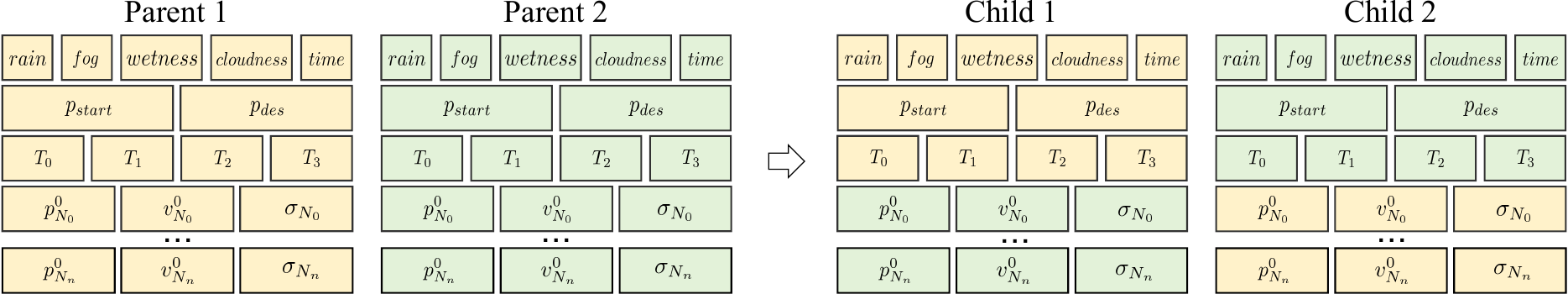}
    \caption{An Example of Crossover}
    \label{img:crossover}
\end{figure}

\textbf{Crossover.} \revision{Crossover introduces inter-individual variation through single-point crossover on chromosomes between two individuals. As shown in Fig.~\ref{img:crossover}, a random crossover point is selected, and the chromosome segments after that point are exchanged to generate new offspring.}

\textbf{Selection.} 
\revision{For seed selection, \tool reuses the energy-based evaluation and selection framework of BehAVExplor~\cite{cheng2023behavexplor}, but adapts it to our scenario representation and testing objective. In BehAVExplor, the energy mechanism is used to balance violation feedback and behavior diversity during seed selection. In \tool, we use the same mechanism to prioritize scenario configurations that are more likely to expose EIVs.}

\revision{\emph{Violation Feedback.} We calculate the violation score $v_I$ of a seed scenario configuration $I$ by Eq.~\ref{eq:violation_feedback},}
\begin{equation}\label{eq:violation_feedback}
    v_I = \sum_{v \in \{collision, Lines, P_{des} \}} f_v(I)
\end{equation}
\revision{where $I$ denotes an executed scenario configuration, and $f_v(I)$ is the violation feedback function for the corresponding specification. Specifically, $f_{collision}(I)$ returns the minimum distance between the Ego vehicle and NPC vehicles during the execution of $I$; once an actual collision occurs, the value becomes zero. Similarly, $f_{Lines}(I)$ returns the minimum distance between the Ego vehicle and illegal lines $Lines$; once the Ego vehicle hits an illegal line, the value becomes zero. Following~\cite{cheng2023behavexplor}, we define $f_{P_{des}}(I)$ as $f_{P_{des}}(I)=\max(10-D_{E2des}(p_E^{n}, p_{des}), 0)$, where $D_{E2des}(p_E^{n}, p_{des})$ denotes the distance between the final position of the Ego vehicle $p_E^{n}$ and its destination $p_{des}$ after simulation. This definition means that once the Ego vehicle is still more than 10 meters away from the destination at the end of the simulation, the corresponding violation feedback becomes zero, indicating that the execution is close to violating the destination-reaching requirement. Therefore, lower values of $v_I$ generally indicate that the corresponding scenario configuration is closer to exposing one or more violations. We use the unweighted sum of these three terms as a unified violation-feedback objective for seed evaluation during search, rather than to model the relative severity of different violations.}

\revision{\emph{Behavior Diversity.}} We calculate a minimum distance $d_{I'}$ to measure the diversity of a new mutant $I'$ by Eq.~\ref{eq:diversity},
\begin{equation}\label{eq:diversity}
    d_{I'} =  \min_{i \in Q} dis(\mathbf{H}_{I'}, \mathbf{H}_i)
\end{equation} 
where \revision{$Q$ is the current seed corpus, $\mathbf{H}_I$ denotes the behavior feature representation of the Ego vehicle in seed $I$, extracted from the observed Ego trajectory after executing $I$, and $\mathbf{H}_i$ denotes the behavior feature representation of the Ego vehicle in seed $i \in Q$. The function $dis(\mathbf{H}_{I'}, \mathbf{H}_i)$ returns the Hamming distance~\cite{hamming1986coding}, which measures the behavioral diversity of the Ego vehicle between two seeds. The larger $d_{I'}$ is, the more diverse the Ego behavior in $I'$.}

\revision{\emph{Energy-Based Selection.}} When the fuzzer starts, all initial seeds generated randomly are assigned with the same energy (\ie 1). \revision{Then, we update the energy $E_I$ of seed scenario configuration $I$ and calculate the energy $E_{I'}$ for the newly mutated scenario configuration $I'$ during fuzzing, where $I'$ is mutated from $I$.} 

Following~\cite{cheng2023behavexplor}, we update the energy of $I$ by Eq.~\ref{eq:energy_update},
\begin{equation}\label{eq:energy_update}
    E_I = E_I + w_1 \cdot \Delta E_F + w_2 \cdot \Delta E_V + w_3 \cdot \Delta E_S
\end{equation}
where $w_1$, $w_2$ and $w_3$ are parameters that can adjust the weights of the three factors. We use the same values as~\cite{cheng2023behavexplor} to set the parameters. $\Delta E_F$ represents the failure frequency. For the seed $I$, we record the number of failed/non-failed test cases (denoted as \#F and \#NF) that are mutated from $I$, and the $\Delta E_F$ is calculated by Eq.~\ref{eq:failure_frequency},
\begin{equation}\label{eq:failure_frequency}
\Delta E_F =
\begin{cases}
\text{\#F} / \text{\#NF} + \text{\#F}, & I' \text{ is a failed test case} \\
-\text{\#NF} \cdot 0.1 / (\text{\#NF} + \text{\#F}), & I' \text{ is a benign test case}
\end{cases} 
\end{equation}
$\Delta E_V$ represents \revision{the violation degree change. As it may be hard to directly generate failed test cases, we consider the violation degree change between the seed $I$ and the newly mutated scenario configuration $I'$}, which is calculated by Eq.~\ref{eq:violation_change}, 
\begin{equation}\label{eq:violation_change}
\Delta E_V = \frac{v_I - v_{I'}}{1 - d_{I'} + \gamma}
\end{equation}
where $\gamma$ is a small value (\eg $10^{-5}$) to avoid zero denominator. $\Delta E_S$ is the selection frequency, which is a fixed energy decay value following~\cite{cheng2023behavexplor}, \ie $\Delta E_S$ = -0.05.

\revision{With respect to the newly generated $I'$, we assign the initial energy $E_{I'}$ by Eq.~\ref{eq:initial_energy}, where $\Delta E_V$ is the violation degree change defined in Eq.~\ref{eq:violation_change}, and $w_2$ is its corresponding weight in Eq.~\ref{eq:energy_update}.}
\begin{equation}\label{eq:initial_energy}
    E_{I'} = 1 + w_2 \cdot \Delta E_V
\end{equation}

Based on the energy, we compute a selection probability of each seed $I$ by Eq.~\ref{eq:selection_probability}, 
\begin{equation}\label{eq:selection_probability}
p_I = \frac{\max(E_I, 0)}{\sum_{i \in Q} E_i}
\end{equation}
\revision{where $Q$ is the current seed corpus, $E_I$ is the energy of seed $I$, and $E_i$ is the energy of seed $i$ in $Q$. We use the max function to make sure all probabilities are nonnegative. Finally, the seed with higher energy will have a higher probability to be selected for breeding the next generation.}


\subsection{Scenario Executor}\label{sec:simulation_executor}

Since the scenario configuration generated in Sec.~\ref{sec:scenario_generator} does not include the NPC vehicle trajectories, we cannot directly utilize the scenario executor from previous works~\cite{li2020av, Tian2022, tian2022mosat, kim2022drivefuzz, zhong2022neural,cheng2023behavexplor, Haq2020}. Moreover, each NPC vehicle in the scenario requires asynchronous behavior generation. We implement a new scenario executor capable of generating NPC vehicle behaviors, recording simulation data, and reporting violation scenarios. We illustrate the process of scenario execution and test oracles in \tool.

\textbf{Execution Process.}
The process of the scenario execution is represented in Algorithm~\ref{alg:simulator_execution}. The input of scenario executor includes the map $Map$, the maximum allowed frame duration $t^S$, and the configuration of scenario $Conf$. The output of scenario executor is the record of the simulation $Record$.

\begin{algorithm}[!t] 
    \caption{Scenario Execution}
    \small
    \label{alg:simulator_execution}
    \LinesNumbered 
    \SetKwFunction{FMonitorSignal}{MonitorSignal} 
    \SetKwProg{Fn}{Function}{:}{}
    \KwIn{the map: $Map$, the maximum frame duration: $t^{S}$, and the configuration of scenario: $Conf$ }
    \KwOut{the record of the simulation: $Record$}

    $NPClist \leftarrow InitNPC(Conf.\ \mathbb{N})$\; \label{init.begin}\label{NPC_init}
    $sim \leftarrow Initialize(Map, Conf, NPClist)$\;\label{init}

    \ForEach{$NPC \in NPClist$}{ \label{set.begin}
        $NPC.\ status \leftarrow  IDLE$\;
    }   \label{init.end}

    \For{$t \leftarrow 1$ \KwTo $t^{S}$}{ \label{run.begin}
        \ForEach{$NPC \in NPClist$}{
            \If {$NPC.\ status = IDLE$}{
                $m \leftarrow NPC.\ DecideManeuver(sim.\ Ego)$\; \label{decide}
                $m.trajectory \leftarrow NPC.\ PlanTrajectory(m, sim.\ Ego)$\; \label{trajectory}
                $NPC.\ status \leftarrow RUNNING$\; \label{status}
                $MonitorSignal(NPC,m)$\; \label{monitor}
            }
        }
        $sim.run(0.1)$\; \label{0.1}
    } \label{run.end}
    $Record \leftarrow UpdateRecord(sim.\ Ego, NPClist)$\; \label{record}
    \Return{Record}\; \label{return}

    \tcp{Asynchronously monitor the execution status of maneuver.}
    \Fn{\FMonitorSignal{$NPC, m$}}{ \label{fun.begin}
        \If{$m.\ execute() = SUCCESS$}{ \label{if_execute}
            $NPC.\ status \leftarrow IDLE$\; \label{mtoIDLE}
        }
    } \label{fun.end}
\end{algorithm}

First, we initialize the scenario (Line~\ref{init.begin}-\ref{init.end}). Specifically, we instantiate the NPC vehicles according to $Conf.\ \mathbb{N}$ (Line~\ref{NPC_init}). We load the map as well as NPC vehicles into the simulator $sim$, initialize the \revision{weather} and traffic signals, and set the driving task of the Ego vehicle using $Conf$, bridging it with the ADS under test (Line~\ref{init}). For each NPC vehicle, we set its status to be $IDLE$ (Line~\ref{set.begin}-\ref{init.end}).

Second, we run the simulation loop for \revision{a total of $t^S$ frames} (Line~\ref{run.begin}-\ref{run.end}). Each frame represents a simulation time step of 0.1 seconds (Line~\ref{0.1}). For each NPC vehicle, if its status is $IDLE$, the NPC vehicle will detect the behavior of the Ego vehicle in the simulator and decide the maneuver $m$ (Line~\ref{decide}), which is introduced in Sec.~\ref{sec:maneuver_decision}. Then, the NPC vehicle will plan the trajectory of $m$ (Line~\ref{trajectory}), which is introduced in Sec.~\ref{sec:trajectory_generation}, and set its status to $RUNNING$ (Line~\ref{status}). Besides, we start an asynchronous task to execute and monitor $m$ by calling the function $MonitorSignal$ (Line~\ref{monitor}). When the simulation is completed, we update the simulation record (Line~\ref{record}), which includes the trajectories of the Ego vehicle and all the NPC vehicles, and save the  record for reproduction and evaluation (Line~\ref{return}).

For the function $MonitorSignal$, it is an asynchronous function independent of the simulation loop (Line~\ref{fun.begin}-\ref{fun.end}). We execute the trajectory of maneuver $m$ and wait for the $SUCCESS$ signal, which indicates that the NPC vehicle has finished one maneuver (Line~\ref{if_execute}). Then, we set the status of $NPC$ to $IDLE$ (Line~\ref{mtoIDLE}) and start to generate another behavior of $NPC$.

\textbf{Violation Analysis.} The scenario executor will also determine whether there are any violations during the simulation. Based on previous work~\cite{cheng2023behavexplor,huai2023doppelganger}, we consider 4 test oracles to assess the ADS's abilities of collision avoidance, reaching destination, not hitting illegal lines, and obeying traffic signals.

\emph{(1) Collision Avoidance.} This oracle checks if the Ego vehicle collides with \revision{an} NPC vehicle. We use the collision detection function provided by the simulator directly to determine the failure of this oracle. Note that, when the collision occurs and is detected by the simulator's callback function, the simulation will end immediately and return the simulation record before the collision.

\emph{(2) Not Hitting Illegal Lines.} This oracle checks if the Ego vehicle hits the illegal lines (\eg yellow lines or edge lines) during the simulation. Given the waypoints $\langle p_{E}^{0},p_{E}^{1}, \dots, p_{E}^{n} \rangle$ of the Ego vehicle and a set of illegal lines denoted as $Lines$ extracted from the map, the failing condition of this oracle is defined by Eq.~\ref{eq:illegal},
\begin{equation}\label{eq:illegal}
    \min ( \{ D_{E2l}(p_E^t, l) \mid l \in Lines, 0 \leq t \leq n  \}) < threshold 
\end{equation}
where $ D_{E2l}(p_E^t, l) $ calculates the distance between the center of the Ego vehicle and the illegal line $l$ at frame $t$, and $threshold$ is set to half width of the Ego vehicle bounding box in our work. 

\emph{(3) Reaching Destination.} This oracle checks if the Ego vehicle reaches the destination in the given time. Given the waypoints $\langle p_{E}^{0},p_{E}^{1}, \dots, p_{E}^{n} \rangle$ of the Ego vehicle after simulation and its destination $p_{des}$, the failing condition of this oracle is defined by Eq.~\ref{eq:destination},
\begin{equation}\label{eq:destination}
    D_{E2des}(p_{E}^n, p_{des}) > threshold
\end{equation}
where $D_{E2des}(p_{E}^n, p_{des})$ evaluates the final distance of the Ego vehicle to its destination, and $threshold$ is set to half length of the Ego vehicle bounding box in our work.

\emph{(4) Obeying Traffic Signals.} This oracle checks if the Ego vehicle crosses the stop line at a positive speed when the signal is red. Given the waypoints $\langle p_{E}^{0},p_{E}^{1}, \dots, p_{E}^{n} \rangle$ and the speed sequence $\langle v_{E}^{0},v_{E}^{1}, \dots, v_{E}^{n} \rangle$ of the Ego vehicle, the traffic signal configuration $T_j$ of the light in Ego direction, and the stop line $l_{stop}$ associated with the traffic light  extracted from the map, the failing condition of this oracle is defined by Eq.~\ref{eq:signal},
\begin{equation}\label{eq:signal}
    \exists t \in [0, n], signal(T_j, t)=RED \wedge D_{E2l}(p^{t}_{E},l_{stop}) = 0 \wedge v^{t}_{E} > 0
\end{equation}
where $signal(T_j, t)$ returns the color of the traffic light $T_j$ at frame $t$, and $D_{E2l}(p^{t}_{E},l_{stop})$ calculates the distance between the center of the Ego vehicle and the stop line $l_{stop}$ at frame $t$.

Through the above test oracles, we can discover the violation scenarios. Then, we can collect the simulation record and replay the driving task of the Ego vehicle along with the recorded trajectories of NPC vehicles in the same environment to reproduce the violations for further diagnosis.

\section{Evaluation}
To evaluate the effectiveness and efficiency of \tool, we design the following \revision{six} research questions.

\begin{itemize}[leftmargin=*]
    \item \textbf{RQ1}: How effective is \tool in finding Ego-induced violations compared with state-of-the-art approaches?
    \item \textbf{RQ2}: How efficient is \tool in finding Ego-induced violations compared with state-of-the-art approaches?
    \item \textbf{RQ3}: Can \tool generate smoother and more various speed sequences of NPC vehicles compared with state-of-the-art approaches?
    \item \textbf{\revision{RQ4}}: \revision{What are the effects of the genetic algorithm-based scenario generator and the three NPC vehicle driving strategies on the testing results?} 
    \item \textbf{\revision{RQ5}}: \revision{How sensitive is \tool to the key safety threshold parameter in maneuver constraints?}
    \item \textbf{\revision{RQ6}}: \revision{Can \tool be effectively migrated to another simulator in more complex traffic scenarios?}
\end{itemize}

\subsection{Evaluation Setup}\label{sec:evaluationsetup}

\textbf{Target ADS and Simulation Platform.} We choose Baidu Apollo 8.0~\cite{apollo} as our target ADS, which is one of the most representative industrial-grade ADSs with widespread commercialization.~We select LGSVL 2021.3~\cite{lgsvl} as our simulation platform because LGVSL \cite{rong2020lgsvl} offers stable connections with Apollo. Although the remote service of LGSVL is no longer maintained, we use a local version~\cite{sora-svl}. \revision{We also evaluate the portability of \tool using CARLA 0.9.14~\cite{{Dosovitskiy17}}.}

\textbf{Prototype.} We implement a prototype of \tool with 13,657 lines of Python code. Our prototype uses simulator Python APIs~\cite{pythonapi} for scenario execution and violation detection. During the process of simulation, Apollo 8.0 is equipped with a wide range of sensors, including two camera sensors, one GPS, one radar and one LiDAR. All modules of Apollo are turned on, including localization module, perception module, prediction module, routing module, planning module and control module. We choose the SanFrancisco map in the SVL map library which contains various types of road.

\textbf{Baselines.} We compare \tool with four state-of-the-art testing approaches, \ie \textsc{AV-Fuzzer}~\cite{li2020av}, \textsc{AutoFuzz} \cite{zhong2022neural}, CRISCO~\cite{Tian2022}, and BehAVExplor~\cite{cheng2023behavexplor}. \textsc{AV-Fuzzer} uses genetic algorithm~to evolve NPC vehicles' maneuvers and speeds after each execution~to find violation scenarios in highway and urban-way (\ie straight~and curve roads). \textsc{AutoFuzz} leverages neural networks to predict outcomes and adopts a gradient-based algorithm for the mutation of positions and speeds of traffic participants (\eg NPC vehicles and pedestrians). CRISCO extracts influential behavior patterns mined from real traffic trajectories (\eg inD~\cite{bock2020ind}) along with random mutation of traffic participants' speeds to generate testing scenarios after each execution. BehAVExplor is a novel behavior-guided fuzzing approach performing maneuvers mutation and trajectories mutation of NPC vehicles to find diverse violation scenarios. \revision{The open-source implementations of these baselines only provide stable integrations with LGSVL. Therefore, we conduct the main comparative experiments on LGSVL to ensure a fair and reproducible comparison.}

\textbf{Research Question Setup.} For \textbf{RQ1}, \textbf{RQ2} and \textbf{RQ3}, we run \tool and other four approaches for 12 hours, generating driving scenarios respectively. Specifically, we run these approaches in a 4-lane straight road and a crossroad equipped with four traffic lights. Note that, we only run \textsc{AV-Fuzzer} in the four-lane straight road because its open-source version does not support crossroad scenarios and is difficult to migrate. Besides, we set the maximum number of vehicles to 4 in each scenario generated by these approaches, ignoring pedestrians and cyclists, and the maximum simulation duration to 30 seconds for a fair comparison. We verify whether the violations generated by these approaches are induced by the Ego vehicle using \textsc{DiaVio}~\cite{lu2024diavio}, an LLM-powered diagnosis tool, and ACAV~\cite{sun2024acav}, a causality-based analysis tool. These violations are then classified into four groups according to the test oracles they violate (\ie 4 test oracles in Sec.~\ref{sec:simulation_executor}).

For \textbf{RQ1}, we evaluate the effectiveness of \tool~from the following aspects: (1) How many EIVs can be~found? (2) What is the proportion of EIVs among all the~violations? (3) How many unique EIV patterns can be found? \revision{To better interpret the discovered EIVs, we group them into \emph{unique EIV patterns}. A unique EIV pattern is defined as a class of Ego-induced violations that share three characteristics: (1) the same violation type exhibited by the Ego vehicle (e.g., rear-end collision, side collision, red-light violation, or destination-missing behavior), (2) the same number of NPC vehicles involved and similar NPC maneuvers triggered, and (3) the same underlying flawed ADS module or causal factor revealed by diagnosis. Two EIVs are assigned to the same pattern only when these three aspects are consistent; otherwise, they are treated as different patterns.} \revision{The grouping procedure is as follows. We first identify whether a reported violation is induced by the Ego vehicle using \textsc{DiaVio}~\cite{lu2024diavio}. For collision violations, we further use ACAV~\cite{sun2024acav} to analyze the causal chain and identify the faulty ADS module. Since ACAV currently only supports collision analysis, the remaining EIVs are further examined through manual diagnosis based on simulator records, vehicle trajectories, traffic-signal states, and the interaction process between the Ego vehicle and NPC vehicles. After that, we assign each confirmed EIV to a pattern according to the above three criteria.} \revision{By abstracting a large number of concrete violation scenarios into a small set of recurring and reproducible failure modes, unique EIV patterns provide a more meaningful way to assess whether a testing approach can uncover diverse violation scenarios rather than repeatedly generating homogeneous ones, while also helping developers more efficiently localize the potential flaws in ADSs.}

For \textbf{RQ2}, we compare \tool with other four approaches from two aspects: (1) How much time does it take to find the first EIV~and how much time does it take to find one EIV on average? (2) What are the proportions of execution time and analysis time in 12 hours? Specifically, execution time is the time taken to run the scenarios, while analysis time includes processes such as environment initialization, feedback collection, and other computing~processes. \revision{We also report the average number of NPC maneuver switches triggered in the simulation.}

For \textbf{RQ3}, we calculate and report the average Standard Deviation of Speed Changes (SDSC) and the number of change points to measure the smoothness and variation of NPC vehicle speed sequences using PELT~\cite{killick2012optimal} algorithm.

For \textbf{RQ4}, \revision{we implement a random approach denoted as \textsc{Rand}, which performs online NPC behavior generation but randomly generates scenario initial configurations. Besides,} we create three variants of \tool, \ie \toolcon, \tooladv and \toolagg, which generate scenarios where all NPC vehicles would only adopt yielding, adversarial, or overtaking driving strategy throughout the simulation, respectively. We run these variants and evaluate them from the aspects concerned in \textbf{RQ1}.

\revision{For \textbf{RQ5}, we study the sensitivity of \tool to the key safety threshold parameter in maneuver constraints regarding acceleration, deceleration, and lane changing maneuvers. By varying the threshold parameter from 20 meters to 40 meters, we analyze whether the effectiveness and efficiency of \tool remain stable under different settings.}

\revision{For \textbf{RQ6}, we migrate \tool to CARLA 0.9.14~\cite{{Dosovitskiy17}} using the official map named Town03, which is a larger, urban map with a roundabout and large junctions. We set the maximum number of vehicles to 8 in each scenario generated by \tool, and the maximum simulation duration to 30 seconds. We report the average results of the effectiveness and efficiency of \tool in the new environment.}

We run all the above experiments 5 times in case of~randomness, applying Mann-Whitney-U-Test~\cite{mcknight2010mann} and computing the Cohen's d~\cite{cohen2013statistical} to evaluate the statistical significance and effect size. We replay the violation scenarios to ensure the reproductivity of the failures, and finally report the average results.

\textbf{Experiment Environment.} We conduct all the experiments on an Ubuntu 22.04.4 LTS server with an NVIDIA GeForce RTX 4090 GPU, Intel Core i9-13900K (32) CPU with 5.500GHz processor and 64GB memory.

\subsection{RQ1: Effectiveness Evaluation}\label{RQ1}
We first present the overall effectiveness of \tool and then discuss the unique EIV patterns we found in detail.

\begin{table}[!t]
    \caption{Results of the General Effectiveness}
    \vspace{-5pt}
    \label{rq1:table1}
    \centering
    \begin{adjustbox}{width=0.75\linewidth}
    \begin{threeparttable}

    \begin{tabular}{ccccccc}
        \toprule
        \makecell{\textbf{Road}} & \textbf{Approach} &\textbf{\#Scenario$^1$} & \textbf{\#Violation} & \textbf{\#EIV}$^2$ & \textbf{Proportion} &\textbf{\#UniP}$^3$\\
        \midrule
        \multirow{6}{*}{\makecell{\textbf{Straight} \\ \textbf{Road}}} &   \textbf{\textsc{AV-Fuzzer}}&665.8 & 245.8&   \cellcolor{gray!15}36.6&   \cellcolor{gray!15}14.89\% & \cellcolor{gray!15}4.8\\
            &   \textbf{\textsc{AutoFuzz}}&893.4& 160.6&   \cellcolor{gray!15}27.6 &  \cellcolor{gray!15}17.19\% & \cellcolor{gray!15}3.0\\
            &   \textbf{CRISCO}&913.6&    138.4&   \cellcolor{gray!15}67.4&  \cellcolor{gray!15}48.70\%& \cellcolor{gray!15}6.6\\
            &   \textbf{BehAVExplor}&645.4&   101.8&    \cellcolor{gray!15}75.8 &   \cellcolor{gray!15}74.46\%& \cellcolor{gray!15}10.2\\
            &   \textbf{\tool}&770.6 &190.2&   \cellcolor{gray!15}\textbf{153.4}&   \cellcolor{gray!15}80.65\%& \cellcolor{gray!15}\textbf{11.8}\\
        \midrule
        \multirow{5}{*}{\textbf{Crossroad}} &   \textbf{\textsc{AutoFuzz}}&944.8 & 112.4&   \cellcolor{gray!15}15.2&   \cellcolor{gray!15}13.52\% & \cellcolor{gray!15}2.6\\
            &   \textbf{CRISCO}&973.2 & 104.0&   \cellcolor{gray!15}33.4&   \cellcolor{gray!15}32.12\% & \cellcolor{gray!15}3.2\\
            &   \textbf{BehAVExplor}&611.8 & 84.8&   \cellcolor{gray!15}47.4 &   \cellcolor{gray!15}55.90\% & \cellcolor{gray!15}4.4\\
            &   \textbf{\tool}&643.0 & 85.0&   \cellcolor{gray!15}\textbf{69.8}&   \cellcolor{gray!15}\textbf{82.12\%} & \cellcolor{gray!15}\textbf{7.2}\\
        \bottomrule
    \end{tabular}
    \begin{tablenotes}
        \footnotesize
        \item $^1$ \textit{the number of the generated scenarios}.
        \item $^2$ \textit{the number of violations induced by the Ego vehicle}.
        \item $^3$ \textit{the number of unique EIV patterns}.
      \end{tablenotes}
\end{threeparttable}
\end{adjustbox}
\end{table}

\textbf{Overall Results.}
For each 12-hour run, on average, 770.6 scenarios and 643.0 scenarios are generated by \tool on the straight road and the crossroad, respectively. After \revision{manual} verification, Table~\ref{rq1:table1} presents the general effectiveness of \tool in finding EIVs compared with other approaches. In terms of the number of EIVs, \tool surpasses the other \todo{four} approaches by \todo{102.37\%} at least on the straight road, and by \todo{47.26\%} at least on the crossroad. 
\revision{In terms of the proportion of EIVs among all reported violations, \tool achieves the best performance at \todo{80.65\% (153.4/190.2)} in the straight road scenarios, whereas \textsc{AV-Fuzzer} yields the lowest proportion at only \todo{14.89\% (36.6/245.8)}.} For the crossroad scenarios, \tool outperforms other approaches, achieving an average of \todo{82.12\% (69.8/85.0)}, while \textsc{AutoFuzz} performs the worst with only \todo{13.52\% (15.2/112.4)}. Overall, the proportion of the violations induced by the Ego vehicle increases by \todo{125.21\%} on average compared with other approaches. In terms of the number of unique EIV patterns, \tool identifies \todo{11.8} and \todo{7.2} patterns on the straight road and crossroad, respectively, outperforming the best results of other approaches, which achieve \todo{10.2} and \todo{4.4} (both by \textsc{BehAVExplor}). \tool improves the number of discovered unique EIV patterns, on average, by at least \todo{39.66\%}.

\revision{\textbf{Breakdown Analysis.}} 
\revision{We further analyze why the baseline approaches exhibit much higher false-positive rates than \tool, and why \tool still cannot eliminate false positives completely. The higher false-positive rates of the baseline approaches mainly stem from their pre-execution generation of NPC behaviors. \textsc{AV-Fuzzer} mutates predefined maneuvers and speeds arbitrarily, which can easily produce overly aggressive interactions. \textsc{AutoFuzz} improves search efficiency neural-guided mutation without ensuring behavior-level reasonableness at runtime, leading to many NPC vehicles appearing in unreasonable locations. \textsc{CRISCO} leverages behavior patterns mined from real trajectories, yet these patterns are still instantiated before execution and cannot adapt to the Ego vehicle's real-time responses. \textsc{BehAVExplor} emphasizes behavior diversity during exploration, but does not explicitly constrain whether NPC decisions remain reasonable under instantaneous signal. As a result, violations reported by these baselines are more likely to be dominated by unreasonable NPC behaviors, leading to higher false-positive rates. The most common violation scenarios generated by these baseline approaches with unreasonable NPC behaviors are shown in Fig.~\ref{NPC Fault}.}

\revision{In contrast, \tool reduces such false positives by generating NPC maneuvers and trajectories online according to the Ego vehicle's runtime behavior and traffic signals, thereby making NPC behaviors more interaction-aware and behaviorally constrained. Nevertheless, \tool still cannot completely eliminate false positives, because its rationality guarantees are implemented through a finite set of maneuver constraints with safety thresholds rather than a complete liability model of realistic driving. Consequently, some borderline cases may still satisfy the designed constraints while remaining overly harsh or effectively unavoidable for the Ego vehicle, especially when the NPC vehicles need to perform lane changing maneuvers. The current maneuver constraints use a safety threshold of 30 meters by default. Although this threshold filters out many clearly unreasonable cut-ins, under the aggressive driving strategy, the speed profile generated by the $s$-$t$ graph may still produce borderline interactions in which the NPC's lane change leaves the Ego vehicle with insufficient time to avoid a rear-end collision. Such cases satisfy the implemented constraints but may still be practically unavoidable for the Ego vehicle, and are therefore counted as false positives. As further discussed in \textbf{RQ5}, increasing the safety threshold can reduce such false positives by making NPC behaviors more conservative.}

To sum up, \tool effectively identifies more EIVs in a given time. This success is attributed to its dynamic generation of NPC vehicle maneuvers and trajectories during each simulation execution. By regulating the rationality of NPC behaviors, \tool minimizes unreasonable behaviors of NPC vehicles, thereby reducing the false positive rate of reported violations. Moreover, \tool introduces mutations to NPC driving strategies, ensuring behavioral diversity throughout the simulation.  
In contrast, other approaches overly pursue finding more violations (\eg collision scenarios or hitting illegal line scenarios) and ignore the rationality of NPC vehicle behaviors (\eg abrupt speed changes or non-compliance with traffic signals), resulting in a high false positive rate of reported violations.

\begin{table}[!t]
    \caption{Results of the Number of Unique EIV Patterns}
    \vspace{-5pt}
    \label{rq1:table2}
    \centering
    \begin{adjustbox}{width=0.75\linewidth}
    \begin{threeparttable}
    \begin{tabular}{cccccc|c}
        \toprule
        \textbf{Approach} & \textbf{\#R1}$^1$ & \textbf{\#R2}$^2$ & \textbf{\#R3}$^3$ & \textbf{\#R4}$^4$& \textbf{Sum} & \textbf{Details}\\
        \midrule
        \textbf{\textsc{AV-Fuzzer}}& 5&   0&   0 &0 &  \cellcolor{gray!15}5&R1-1\textasciitilde R1-5    \\
        \textbf{\textsc{AutoFuzz}}& 4&   0&  2&0 & \cellcolor{gray!15}6&R1-1, R1-2,  R1-8, R1-9, R3-3, R3-4   \\
        \textbf{CRISCO}&    7&   2&  2&0 &  \cellcolor{gray!15}11& R1-1\textasciitilde R1-3, R1-6, R1-8\textasciitilde R1-10, R2-1, R2-3, R3-3, R3-4    \\
        \textbf{BehAVExplor}&   9&    3&   4& 0 & \cellcolor{gray!15}16& R-1\textasciitilde R1-5, R1-7, R1-8\textasciitilde R1-10,
        R2-1\textasciitilde R2-3, R3-1\textasciitilde R3-4   \\
        \textbf{\tool}& \textbf{12}&   \textbf{3}&   \textbf{4}&  \textbf{1} &  \cellcolor{gray!15}\textbf{20} & R1-1\textasciitilde R1-12, R2-1\textasciitilde R2-3, R3-1\textasciitilde R3-4, R4-1 \\
        \bottomrule
    \end{tabular}
    \begin{tablenotes}
        \footnotesize
        \item $^1$ \textit{the number of patterns where the Ego vehicle collides with NPC vehicles}.
        \item $^2$ \textit{the number of patterns where the Ego vehicle hits illegal lines}.
        \item $^3$ \textit{the number of patterns where the Ego vehicle gets stuck that fails to reach the destination}.
        \item $^4$ \textit{the number of patterns where the Ego vehicle runs red lights}.
      \end{tablenotes}
\end{threeparttable}
    \end{adjustbox}
\end{table}

\begin{figure*}[!t]
    \centering
    \includegraphics[width=\textwidth]{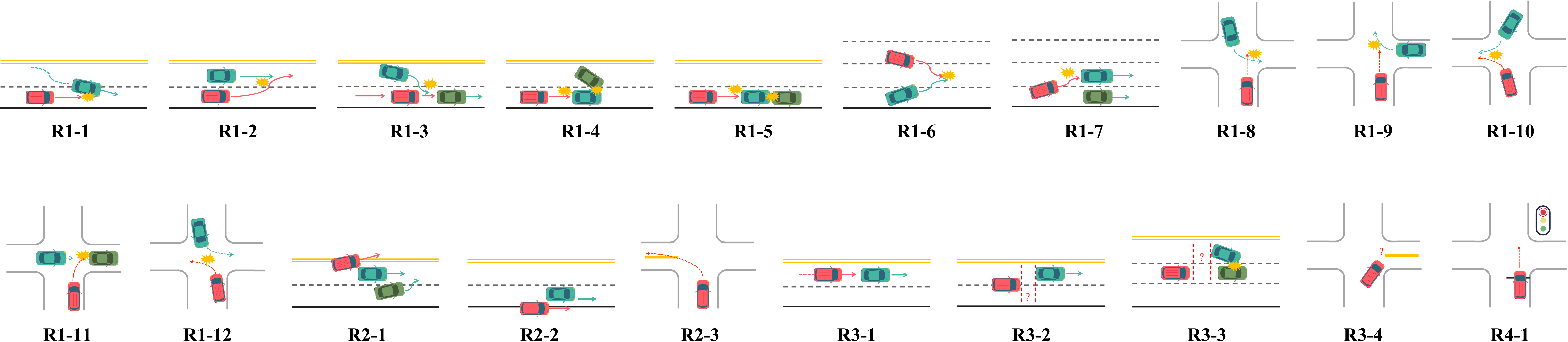}
    \caption{Overview of Each Ego-induced Violation Pattern}
    \label{fig:rq1}
\end{figure*}

\textbf{Unique EIV Patterns.}
Table~\ref{rq1:table2} lists the total number of unique EIV patterns that violate different test oracles in all 5 repetitions of experiments in these 2 types of road (\ie straight road and crossroad). \tool finds 20 unique EIV patterns in total with 12 patterns where the Ego vehicle collides with NPC vehicles, 3 patterns where the Ego vehicle hits illegal lines, 4 patterns where the Ego vehicle gets stuck and fails to reach the destination and 1 pattern where the Ego vehicle runs red lights. The result shows that \tool can discover the most number of unique EIV patterns and cover all unique EIV patterns found by baselines. Fig.~\ref{fig:rq1} shows an overview of each EIV pattern, where the red car denotes the Ego vehicle and green cars represent NPC vehicles. 
We provide an in-depth discussion for each of these unique EIV patterns as follows.

\textbf{Case Study R1-1.}
When the NPC vehicle initiates a lane change, it satisfies all behavior constraints, including maintaining a safe longitudinal distance, activating the turn signal, and performing the maneuver in a non-solid-line area. \revision{Finally, the NPC vehicle finishes the lane change, however, the Ego vehicle fails to decelerate in time, leading to a collision. This case shows a potential issue in the prediction module, as the Ego vehicle has not accurately anticipated the speed of NPC vehicle's lane changing behavior.}

\textbf{Case Study R1-2.}
When the Ego vehicle changes lanes, it fails to maintain a safe longitudinal distance and avoid the NPC vehicle approaching from behind in the target lane, resulting in a collision. According to the root cause analysis, this case shows potential issues in the prediction and planning module, as the Ego vehicle have miscalculated the speed and position of the NPC vehicle and give a wrong driving decision.

\textbf{Case Study R1-3.}
The NPC vehicle performs a compliant lane change \revision{with a low speed}, activating the turn signal and keeping a sufficient distance before merging. The Ego vehicle \revision{chooses to accelerate forward, but ignores another NPC vehicle that is decelerating about 30 meters ahead, ultimately leading to a collision. This case shows that the Ego vehicle lacks effective response to lane changing behavior and risk assessment, with flaw in the planning module.}

\textbf{Case Study R1-4.}
Two NPC vehicles collide during a lane change and come to a stop ahead. After the collision, both NPCs stop far from the Ego vehicle and activate their brake lights. According to the root cause analysis, the Ego vehicle \revision{incorrectly perceives the two stationary vehicles as a single large obstacle in the adjacent lane, while losing the perception of the NPC vehicle blocking its current lane,} and fails to stop in time. This suggests that the Ego vehicle has limitations in \revision{the perception module, which is responsible for object detection and classification errors}.

\textbf{Case Study R1-5.}
The Ego vehicle is following two NPC vehicles. When the two NPC vehicles collide and stop abruptly, the Ego vehicle fails to react in time \revision{because the bounding boxes of the two stopped vehicles overlap in the perception results, making it difficult to accurately estimate the safe distance to the obstacles ahead,} resulting in a collision with the NPC vehicles. This case suggests a lack of effective obstacle perception and \revision{safe distance estimation}.

\textbf{Case Study R1-6.}
The Ego vehicle and an NPC vehicle attempt to change into the same lane simultaneously. Due to the weak lateral awareness and poor coordination, the Ego vehicle fails to detect the NPC vehicle's movement according to root cause analysis, resulting in a side collision during the merging process.

\textbf{Case Study R1-7.}
The Ego vehicle is following a slow-moving NPC vehicle and decides to change lanes to overtake. However, the Ego vehicle fails to avoid another NPC vehicle in the adjacent lane and collides with it during the lane change, indicating problems in the perception and planning modules according to root cause analysis.

\textbf{Case Study R1-8.}
The Ego vehicle is going straight through an intersection when an NPC vehicle from the opposite direction makes a left turn. Since the NPC vehicle is moving slowly, the Ego vehicle marks it with a \textit{Ignored} tag by wrong priority prediction and fails to properly predict its turning behavior, resulting in a side collision. 

\textbf{Case Study R1-9.}
The Ego vehicle is going straight through an intersection when an NPC vehicle from the opposite direction makes a right turn across its path. Due to the slow speed of the NPC vehicle, the Ego vehicle tags it as \textit{Ignored} by wrong priority prediction and does not yield or decelerate, colliding on the side of the NPC vehicle.

\textbf{Case Study R1-10.}
The Ego vehicle is making a left turn at an intersection while an NPC vehicle from the opposite direction is making a right turn. The NPC vehicle is moving slowly, leading the Ego vehicle to assign it an \textit{Ignored} tag by wrong priority prediction and proceed without sufficient caution, resulting in a side collision. 

\textbf{Case Study R1-11.}
When the Ego vehicle turns right to merge into traffic, it predicts the speed of the slowly moving vehicle behind a fast vehicle wrongly, and assigns the slow vehicle an \textit{Ignored} tag. Consequently, the Ego vehicle collides with the NPC vehicle while turning, showing a problem in the prediction module.

\textbf{Case Study R1-12.}
The Ego vehicle and an NPC vehicle simultaneously turn left from opposite directions at an intersection. Due to inaccurate prediction of the NPC vehicle's trajectory and the shaking of the Ego vehicle when turning, a side collision occurs. Root cause analysis indicates that there are problems in the prediction module and the control module when handling intersection turning behavior.

\textbf{Case Study R2-1.}
The Ego vehicle changes lanes across the yellow line when overtaking if the NPC vehicles block the feasible lanes, indicating a problem in the planning module.

\textbf{Case Study R2-2.}
The Ego vehicle moves on the road boundaries to perform the overtaking behavior, indicating a problem in the planning module.

\textbf{Case Study R2-3.}
In a left-turn scenario, if the speed of the Ego vehicle is slow and the steering angle is relatively large, the Ego vehicle will hit the yellow line and continue to press the yellow line after left turn, indicating a problem in the control module.

\textbf{Case Study R3-1.}   
The Ego vehicle follows a slow-moving NPC vehicle without an overtaking maneuver, despite having sufficient conditions to do so. As a result, the Ego vehicle fails to reach the destination within the required time limit. This indicates a problem in the planning module regarding overtaking strategies in low-speed following scenarios.

\textbf{Case Study R3-2.}
The Ego vehicle follows an NPC vehicle whose speed is slow, and it hesitates between continuing to follow and initiating a lane change maneuver. As a result, it becomes stuck in decision-making and fails to complete an overtaking. This case reveals limitations in the planning module.

\textbf{Case Study R3-3.}
Two NPC vehicles collide ahead of the Ego vehicle, blocking the lane and forming a static obstacle. The Ego vehicle fails to plan a feasible path to change lanes and bypass the obstacle, resulting in it being stuck behind the blocked lane. This indicates limitations in the routing and planning modules.

\textbf{Case Study R3-4.}
When the turning arc is small, the Ego vehicle may plan off-road trajectories, leading to fluctuating steering behavior. Eventually, the vehicle gets stuck, showing issues in both the planning and control modules.
 
\textbf{Case Study R4-1.} 
The Ego vehicle stops when the signal is red and passes the stop line. Then, it loses the perception of the traffic signal and restarts to run a red light through the intersection, indicating a bug in the perception module.

\begin{tcolorbox}[size=small, opacityfill=1, before skip=5pt, after skip=10pt]
    \textit{\textbf{Summary.}} \tool can not only find more Ego-induced violations compared with other approaches, increasing the proportion of EIVs among all reported violations, on average, by \todo{125.21\%}, but also can discover more unique EIV patterns effectively, improving the number of discovered unique EIV patterns, on average, by at least \todo{39.71\%}.
\end{tcolorbox}

\subsection{RQ2: Efficiency Evaluation}\label{RQ2}

\begin{figure}[!t]
    \centering
    \subfloat[Straight Road]{\includegraphics[width=0.5\linewidth]{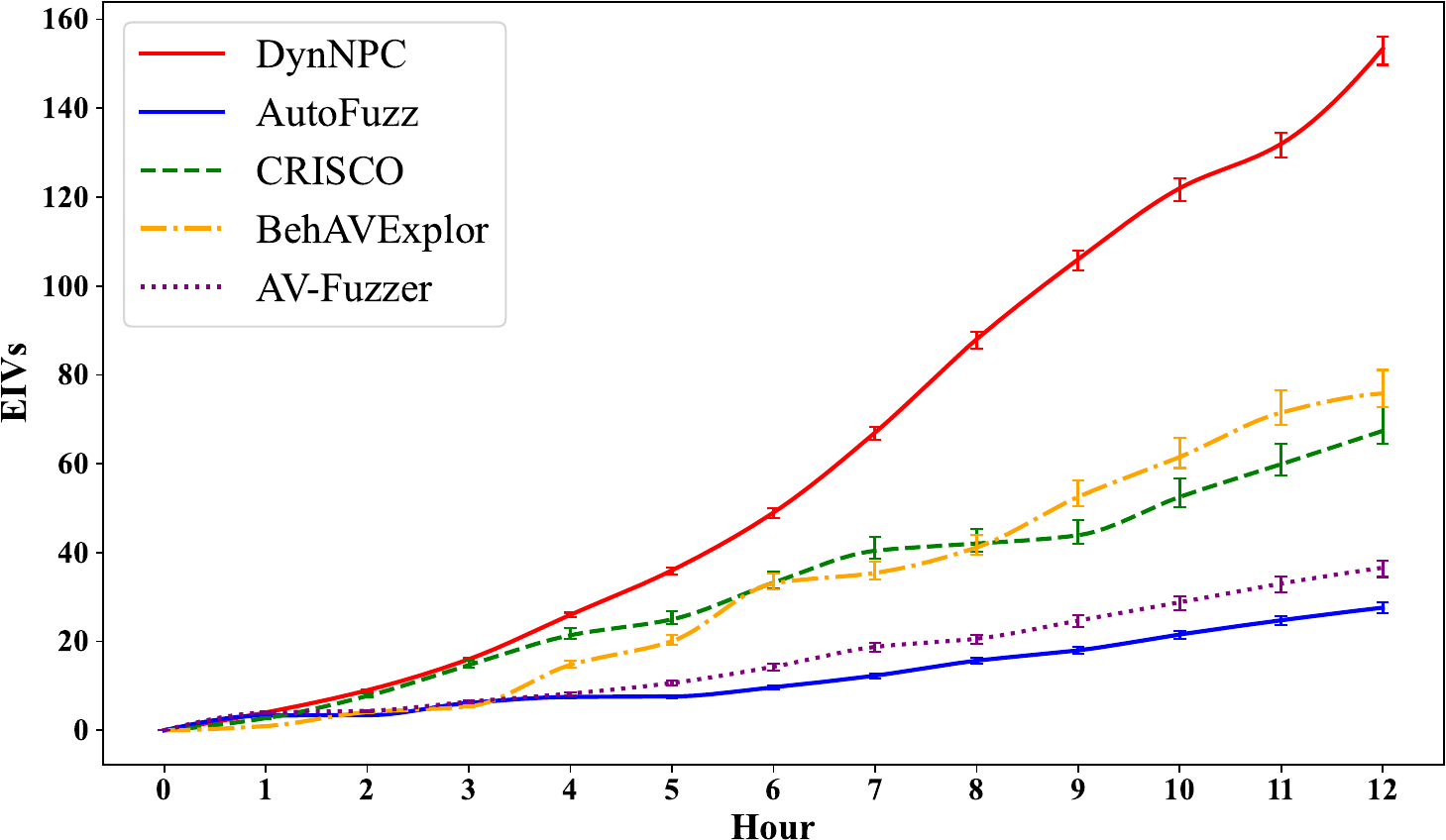}}
    \subfloat[Crossroad]{\includegraphics[width=0.5\linewidth]{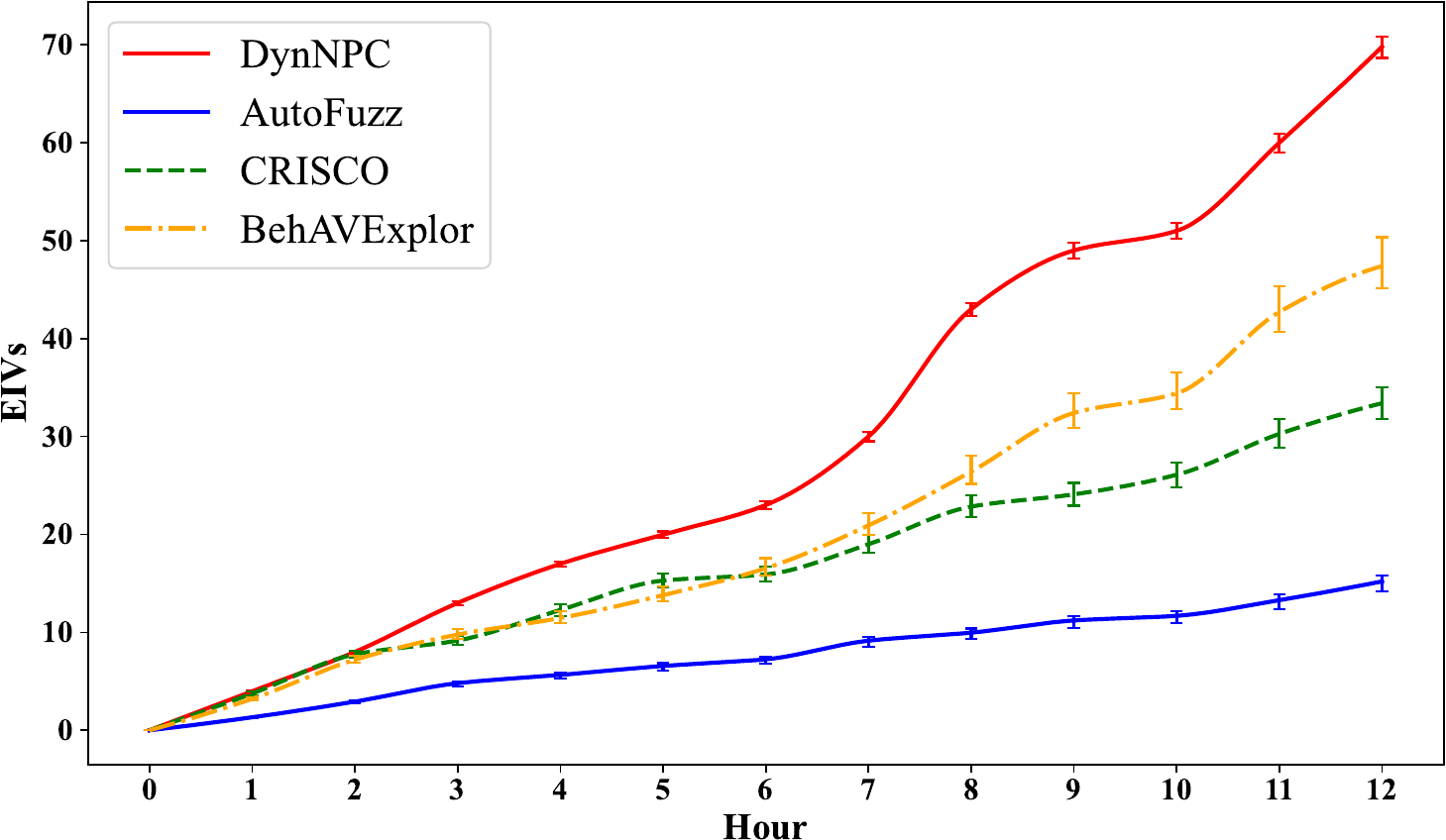}}
    \caption{Results of the General Efficiency}
    \label{rq2:fig1}
\end{figure}

\begin{table}[!t]
    \caption{Efficiency in Finding Ego-induced Violations}
    \vspace{-5pt}
    \label{rq2:table3}
    \centering
    \begin{adjustbox}{width=0.5\linewidth}
    \begin{threeparttable}
    \begin{tabular}{cccc}
        \toprule
        \textbf{Road Type} & \textbf{Approach} & \textbf{First EIV}$^1$ & \textbf{One EIV}$^2$\\
        \midrule
        \multirow{6}{*}{\textbf{Straight Road}} &   \textbf{\textsc{AV-Fuzzer}}& 80.70&   19.31\\
            &   \textbf{\textsc{AutoFuzz}}& 130.15&   25.71\\
            &   \textbf{CRISCO}&    7.40&   10.69\\
            &   \textbf{BehAVExplor}&   31.83&    9.44\\
            &   \textbf{\tool}& \textbf{4.69}&   \textbf{4.67}\\
        \midrule
        \multirow{5}{*}{\textbf{Crossroad}} 
            &   \textbf{\textsc{AutoFuzz}}& 107.87&   44.17\\
            &   \textbf{CRISCO}& \textbf{10.69}&   21.36\\
            &   \textbf{BehAVExplor}& 29.43&   15.22\\
            &   \textbf{\tool}& 15.30&   \textbf{10.14}\\
        \bottomrule
    \end{tabular}
    \begin{tablenotes}
        \footnotesize
        \item $^1$ \textit{the time used to find the first EIV (minute)}.
        \item $^2$ \textit{the average time used to find one EIV (minute)}.
      \end{tablenotes}
    \end{threeparttable}
\end{adjustbox}
\end{table}

Fig.~\ref{rq2:fig1} and Table~\ref{rq2:table3} presents the efficiency of \tool in finding EIVs. 
For the straight road scenarios, \tool achieves the shortest time to identify the first EIV, requiring only 4.69 minutes, with CRISCO ranking second. In contrast, AutoFuzz exhibits the poorest performance, taking 130.15 minutes. For the crossroad scenarios, CRISCO takes the shortest time (10.69 minutes) and AutoFuzz takes the longest time (107.87 minutes) to find the first EIV, while \tool uses 15.30 minutes. Overall, \tool improves efficiency in finding the first EIV by \todo {92.50\%} on the straight road, and by \todo{68.98\%} on the crossroad, on average, compared with other approaches.

In terms of the average time to find an EIV, \tool consistently ranks first across both straight road and crossroad scenarios among all the approaches, achieving an average time of 7.41 minutes. Notably, although CRISCO identifies EIVs quickly in the early stages by leveraging trajectories from real-world datasets, its later use of random trajectory mutations introduces numerous unreasonable NPC behaviors, resulting in less efficient overall performance.
On average, \tool achieves a reduction in the time to find one EIV ranging from \todo{41.96\%} to \todo{82.94\%}. Results show that the dynamic generation of NPC vehicle behaviors during simulation execution can significantly accelerate the search for~EIVs.

\begin{table}[!t]
    \caption{Results of Performance Characteristics}
    \vspace{-5pt}
    \label{rq2:table4}
    \centering
    \begin{adjustbox}{width=0.7\linewidth}
    \begin{tabular}{cccccc}
        \toprule
        \textbf{Time} & \textbf{\textsc{AV-Fuzzer}} & \textbf{\textsc{AutoFuzz}} & \textbf{CRISCO} & \textbf{BehAVExplor} & \textbf{\tool}\\
        \midrule
        \textbf{Execution}& 10.65&   11.25&   11.30 & 10.82&  11.18 \\
        \textbf{Analysis}& 1.35&   0.75&  0.70& 1.18&   0.82 \\
        \bottomrule
    \end{tabular}
    \end{adjustbox}
\end{table}

In addition, we calculate the average execution and analysis time across all scenarios for each approach. Table~\ref{rq2:table4} shows the results in detail. \tool uses, on average, 0.82 hours for analysis and 11.18 hours for execution during the simulation. \textsc{AutoFuzz} and CRISCO have shorter analysis time than \tool at the cost of the diversity of NPC vehicle trajectories. Specifically, \textsc{AutoFuzz} employs a constrained neural network evolutionary search method to generate scenarios where NPC vehicles seldom change speeds, thereby reducing the complexity of NPC vehicle trajectory computation. CRISCO, on the other hand, constructs NPC vehicle waypoints by randomly mutating the combination of predefined influential behavior patterns extracted from datasets, leading to reduced computational overhead. Compared with \textsc{AV-Fuzzer} and BehAVExplor, \tool reduces the analysis time by \todo{39.26\%} and \todo{30.51\%}, respectively. The analysis time of \textsc{AV-Fuzzer} and BehAVExplor are prolonged due to complex computation of NPC vehicle trajectories before each execution. However, \tool puts this task into parallel during the execution process, reducing the additional analysis time. 
In general, \tool reduces the analysis time by \todo{39.71\%} on average than other approaches.

\revision{To further characterize how dynamically interactive the generated NPC behaviors are in practice, we additionally measure the average number of NPC vehicle maneuver switches triggered during simulation. The results show a clear difference between \tool and most baselines. In \textsc{AutoFuzz}, CRISCO, and BehAVExplor, each NPC vehicle follows a predefined behavior once the scenario starts, and thus the average number of maneuver switches during execution is 0. \textsc{AV-Fuzzer} is the only exception, as its NPC vehicles randomly switch maneuvers (\ie going straight and changing lane) every five seconds, resulting in an average of 3.76 switches on the straight road. In contrast, \tool enables NPC vehicles to adjust their behaviors online according to the Ego vehicle's real-time states and traffic conditions, leading to an average of 3.25 maneuver switches per NPC on the straight road and 1.12 on the crossroad. These results indicate that the NPC vehicles generated by \tool are not limited to executing a single predefined maneuver, but instead exhibit substantial runtime behavioral adaptation. This dynamic interaction is especially frequent on the straight road, where repeated acceleration, deceleration, lane changing and parking decisions can be triggered during a single run, while it is relatively lower at crossroads due to shorter interaction windows and stronger traffic-signal constraints.} \revision{Such runtime maneuver adaptation also helps improve testing efficiency, as it allows NPC vehicles to produce richer and more targeted interactions with the Ego vehicle within a single run, thereby increasing the chance of exposing EIVs without relying on repeated exploration of many pre-defined trajectory variants.}

\begin{tcolorbox}[size=small, opacityfill=1, before skip=5pt, after skip=10pt]
    \textit{\textbf{Summary.}} \tool can find Ego-induced violations more quickly compared with other approaches, reducing the time to find the first EIV and the average time to find one EIV by \todo{82.13\%} and \todo{60.70\%}, respectively. 
    \revision{Besides, \tool maintains competitive runtime overhead while enabling dynamically interactive NPC behaviors, which helps improve testing efficiency by producing richer and more targeted interactions within a single run.}
\end{tcolorbox}


\begin{table}[!t]
    \caption{Smoothness and Variation of Speed Sequences}
    \vspace{-5pt}
    \label{rq3:table5}
    \centering
    \begin{adjustbox}{width=0.5\linewidth}
    \begin{threeparttable}

    \begin{tabular}{cccccc}
        \toprule
        \textbf{Road Type} & \textbf{Trajectory} & \textbf{SDSC}$^1$ & \textbf{\#ChangeP}$^2$\\
        \midrule
        \multirow{6}{*}{\textbf{Straight Road}}  
        &   \textbf{\textsc{AV-Fuzzer}}  & 2.08  & 24.42 \\
            &   \textbf{\textsc{AutoFuzz}}  & 0.18  & 0.43 \\
            &   \textbf{CRISCO}  & 0.01  & 0.00 \\
            &   \textbf{BehAVExplor}  & 0.33  & 3.26 \\
            &   \cellcolor{gray!15}\textbf{\tool}  & \cellcolor{gray!15}0.12  & \cellcolor{gray!15}2.81 \\

        \midrule
        \multirow{5}{*}{\textbf{Crossroad}} 
            &   \textbf{\textsc{AutoFuzz}}  & 0.15  & 1.89 \\
            &   \textbf{CRISCO}  & 0.01  & 0.00 \\
            &   \textbf{BehAVExplor}  & 0.35  & 6.20 \\
            &   \cellcolor{gray!15}\textbf{\tool}  & \cellcolor{gray!15}0.09  & \cellcolor{gray!15}5.70 \\

        \bottomrule
    \end{tabular}
    \begin{tablenotes}
        \footnotesize
        \item $^1$ \textit{the standard deviation of speed changes.}
        \item $^2$ \textit{the number of change points in speed sequences}.
      \end{tablenotes}
\end{threeparttable}
\end{adjustbox}
\end{table}

\subsection{RQ3: Smoothness and Variation of NPC Vehicle Speeds} \label{RQ3}
\begin{figure*}[!t]
    \centering
    \subfloat[AV-Fuzzer]{
        \includegraphics[width=0.1875\linewidth]{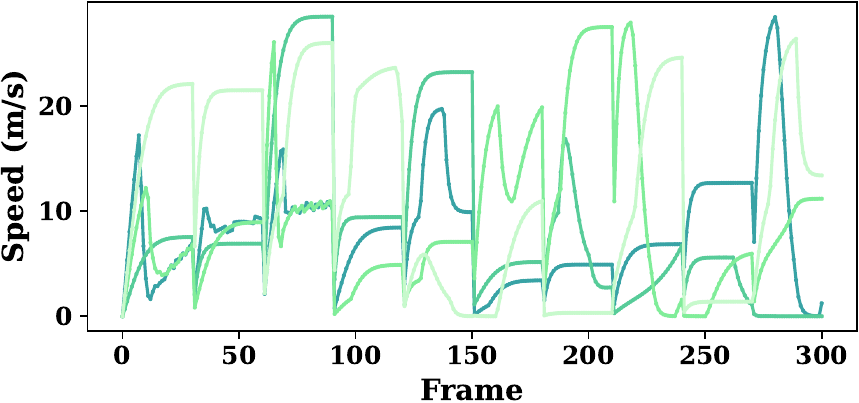}
        \label{fig:AV-Fuzzer}
    }
    \subfloat[AutoFuzz]{
        \includegraphics[width=0.1875\linewidth]{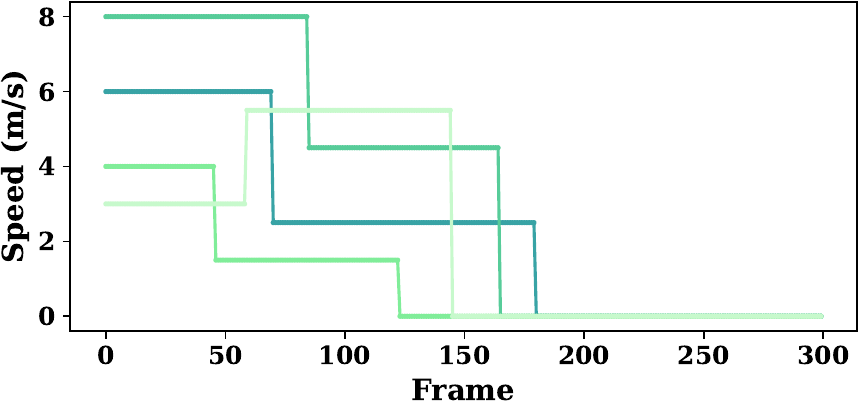}
        \label{fig:AutoFuzz}
    }
    \subfloat[CRISCO]{
        \includegraphics[width=0.1875\linewidth]{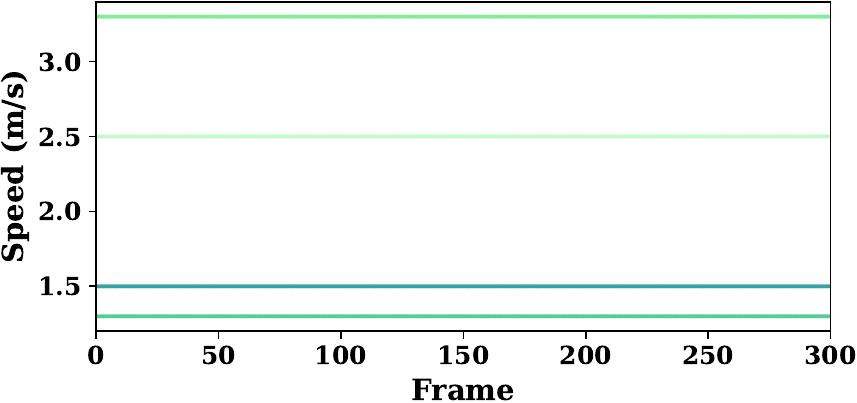}
        \label{fig:CRISCO}
    }
    \subfloat[BehAVExplor]{
        \includegraphics[width=0.1875\linewidth]{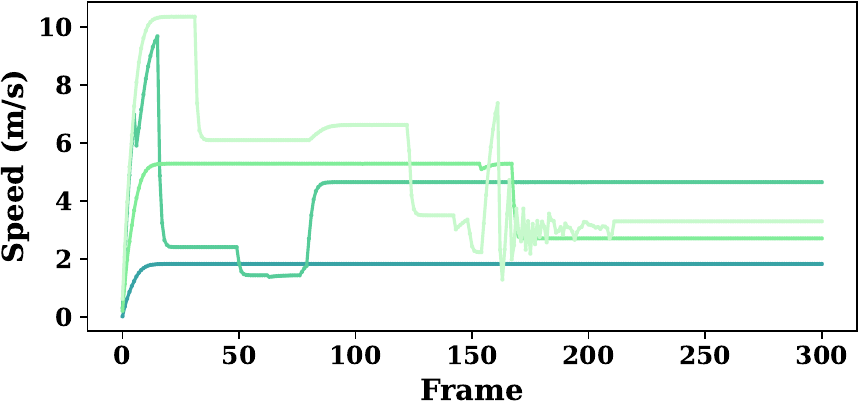}
        \label{fig:BehAVExplor}
    }
    \subfloat[\tool]{
        \includegraphics[width=0.1875\linewidth]{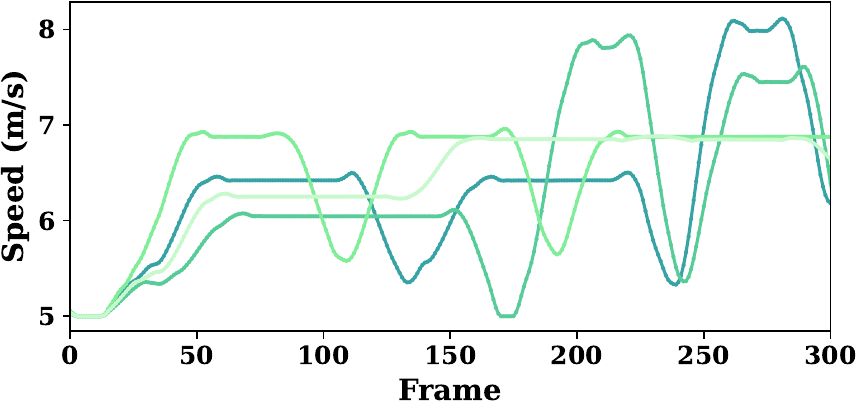}
        \label{fig:dynNPC}
    }
    \caption{Examples of Speed Sequences Generated by Different Approaches.}
    \label{fig:npc_speeds}
\end{figure*}

Table~\ref{rq3:table5} presents the smoothness and variation 
of NPC vehicle speed sequences generated by different approaches within the simulator.
\textsc{AV-Fuzzer} has the highest standard deviation of speed changes (SDSC) and the greatest number of change points (\#ChangeP) in straight road scenarios, indicating that the speed sequences of NPC vehicles generated by \textsc{AV-Fuzzer} exhibit excessive fluctuations and lack smoothness. This is because \textsc{AV-Fuzzer} divides the driving scenario into 5-second time slices and incorporates the behavior of NPC vehicles in each time slice into its genetic representation. During the mutation process, it alters the speed of vehicles within each time slice, leading to significant and abrupt speed changes illustrated by an example of speed sequences~shown~in~Fig.~\ref{fig:AV-Fuzzer}.

\textsc{AutoFuzz} demonstrates low SDSC values both on the straight road and crossroad. However, its number of change points is notably low (0.43 on straight roads and 1.89 on crossroads), indicating that it generates overly simplistic speed profiles with minimal variations. As shown in Fig.~\ref{fig:AutoFuzz}, \textsc{AutoFuzz} adjusts the fixed speed of vehicles over specific time periods in the scenario configurations based on test feedback, ultimately bringing the vehicle to a stop on the map. It fails to avoid occasional abrupt speed transitions and its lack of dynamic changes in speed sequences may fail to capture the complexity of real-world driving behavior.

CRISCO produces nearly constant speed sequences of NPC vehicles, with SDSC values close to zero and no change points in both straight road scenarios and crossroad scenarios. As shown in Fig.~\ref{fig:CRISCO}, the NPC vehicles in CRISCO drive with fixed speeds specified in the scenario configurations. While this approach ensures the maximum smoothness, it fails to simulate diverse driving behaviors, as real-world driving inevitably involves various speed variations.

BehAVExplor performs best among the four baselines, generating NPC vehicle speed sequences with acceptable SDSC values and \#ChangeP values in both straight road scenarios and crossroad scenarios. It constrains the speed of each maneuver taken by NPCs. However, as shown in Fig.~\ref{fig:BehAVExplor}, it still inevitably introduces unreasonable speed changes of NPC vehicles in the scenario configurations due to the guidance of finding collisions as much~as ~possible.

With respect to \tool, on the straight road, it exhibits an SDSC of 0.12 and a \#ChangeP of 2.81, on average. On the crossroad, it achieves an SDSC of 0.09 and a \#ChangeP of 5.70, on average, indicating that our generated speed profiles are not simple constant-speed lines but contain meaningful variations. As an example of the speed sequences generated by \tool, shown in Fig.~\ref{fig:dynNPC}, the speed sequences are smooth and have moderate variations, indicating that \tool can generate speed sequences with better smoothness and speed variations compared with other four approaches.

\begin{tcolorbox}[size=small, opacityfill=1, before skip=5pt, after skip=10pt]
    \textit{\textbf{Summary.}} \tool can generate speed sequences  with better smoothness and variations compared with other four approaches. The planning of speed during simulation execution based on $s \text{-} t$ graph ensures the smoothness and the mutation of driving strategy improves the variation of speed sequences.
\end{tcolorbox}


\subsection{RQ4: \revision{Effects of GA-based Generator and Different Driving Strategies}}\label{RQ4}

\revision{We first evaluate the effect of the GA-based scenario generator by comparing \tool with \textsc{Rand}, and then analyze the effects of different driving strategies using the three variants of \tool.}

\begin{table}[!t]
    \caption{\revision{Effects of GA-based Generator on the General Effectiveness}}
    \vspace{-5pt}
    \label{rq4:table6}
    \centering
    \begin{adjustbox}{width=0.75\linewidth}
    \begin{threeparttable}
    {
    \begin{tabular}{ccccccc}
        \toprule
        \makecell{\textbf{Road}} & \textbf{Approach} &\textbf{\#Scenario$^1$} & \textbf{\#Violation} & \textbf{\#EIV}$^2$ & \textbf{Proportion} &\textbf{\#UniP}$^3$\\
        \midrule
        \multirow{2}{*}{\makecell{\textbf{Straight} \\ \textbf{Road}}} &   \textbf{\textsc{Rand}}&910.8 & 112.0&   \cellcolor{gray!15}88.7&   \cellcolor{gray!15}79.20\% & \cellcolor{gray!15}5.8\\
            &   \textbf{\tool}&770.6 &190.2&   \cellcolor{gray!15}\textbf{153.4}&   \cellcolor{gray!15}\textbf{80.65\%}& \cellcolor{gray!15}\textbf{11.8}\\
        \midrule
        \multirow{2}{*}{\textbf{Crossroad}} &   \textbf{\textsc{Rand}}&794.8 & 66.4&   \cellcolor{gray!15}53.9&   \cellcolor{gray!15}80.11\% & \cellcolor{gray!15}2.6\\

            &   \textbf{\tool}&643.0 & 85.0&   \cellcolor{gray!15}\textbf{69.8}&   \cellcolor{gray!15}\textbf{82.12\%} & \cellcolor{gray!15}\textbf{7.2}\\
        \bottomrule
    \end{tabular}
    }
    \begin{tablenotes}
        \footnotesize
        \item $^1$ \textit{the number of the generated scenarios}.
        \item $^2$ \textit{the number of violations induced by the Ego vehicle}.
        \item $^3$ \textit{the number of unique EIV patterns}.
      \end{tablenotes}
\end{threeparttable}
\end{adjustbox}
\end{table}

\revision{Table~\ref{rq4:table6} presents the effects of the GA-based scenario generator on the general effectiveness. Compared with \textsc{Rand}, \tool consistently finds more EIVs and more unique EIV patterns in both straight road and crossroad scenarios. Specifically, on the straight road, \tool increases the average number of EIVs from 88.7 to 153.4 and the number of unique EIV patterns from 5.8 to 11.8. On the crossroad, \tool improves the average number of EIVs from 53.9 to 69.8 and the number of unique EIV patterns from 2.6 to 7.2. The proportion of EIVs generated by \tool is also slightly higher than that of \textsc{Rand} in both scenarios.}

\revision{Although \textsc{Rand} generates more scenarios within 12 hours, its effectiveness is clearly worse than \tool considering the violation rate. Besides, during five repeated experiments, \textsc{Rand} only find 9 unique EIV patterns (\ie R1-1, R1-2, R1-3, R1-8, R1-9, R2-2, R3-1, R3-3, and R3-4) in 12 hours. 
These results indicate that the initial scenario configurations, such as the initial positions of NPC vehicles, traffic signal settings, and environment conditions, still largely determine whether the Ego vehicle and NPC vehicles can enter meaningful interactions. By adopting genetic algorithm to optimize these high-level scenario configurations, \tool is able to generate fewer but more effective scenarios, thereby exposing substantially more EIVs and diverse EIV patterns. Therefore, the GA-based generator is complementary to runtime NPC behavior generation, rather than redundant with it.}

\begin{table}[!t]
    \caption{Effects of Different Driving Strategies on the General Effectiveness}
    \vspace{-5pt}
    \label{rq4:table7}
    \centering
    \begin{adjustbox}{width=0.75\linewidth}
    \begin{threeparttable}
    \begin{tabular}{ccccccc}
        \toprule
        \makecell{\textbf{Road}} & \textbf{Approach} &\textbf{\#Scenario$^1$} & \textbf{\#Violation} & \textbf{\#EIV$^2$} & \textbf{Proportion} &\textbf{\#UniP$^3$}\\
        \midrule
        \multirow{4}{*}{\makecell{\textbf{Straight} \\ \textbf{Road}}} 
        
            &   \textbf{\toolcon} &767.4 & 228.8&   \cellcolor{gray!15}185.4&  \cellcolor{gray!15}\textbf{81.03\%} & \cellcolor{gray!15}7.8\\
            &   \textbf{\tooladv} & 792.8 &    424.4&   \cellcolor{gray!15}\textbf{291.0}&  \cellcolor{gray!15}68.57\%& \cellcolor{gray!15}10.0\\
            &   \textbf{\toolagg} &734.0 &   161.0&    \cellcolor{gray!15}38.8&   \cellcolor{gray!15}24.10\%& \cellcolor{gray!15}4.4\\
            &   \textbf{\tool}& 770.6 & 190.2&   \cellcolor{gray!15}153.4&   \cellcolor{gray!15}80.65\%& \cellcolor{gray!15}\textbf{11.8}\\

        \midrule
        \multirow{4}{*}{\textbf{Crossroad}} 
        
            &   \textbf{\toolcon} &626.0            & 68.8&   \cellcolor{gray!15}58.4 &   \cellcolor{gray!15}\textbf{84.88\%} & \cellcolor{gray!15}4.8\\
            &   \textbf{\tooladv} &672.4            & 161.0&   \cellcolor{gray!15}\textbf{97.0}&   \cellcolor{gray!15}60.25\% & \cellcolor{gray!15}5.0\\
            &   \textbf{\toolagg} &651.4 & 42.4&   \cellcolor{gray!15}7.0&   \cellcolor{gray!15}16.51\% & \cellcolor{gray!15}1.8\\
            &   \textbf{\tool} & 643.0 & 85.0&   \cellcolor{gray!15}69.8&   \cellcolor{gray!15}82.12\% & \cellcolor{gray!15}\textbf{7.2}\\
        \bottomrule
    \end{tabular}
        \begin{tablenotes}
        \footnotesize
        \item $^1$ \textit{the number of the generated scenarios}.
        \item $^2$ \textit{the number of violations induced by the Ego vehicle}.
        \item $^3$ \textit{the number of unique EIV patterns}.
      \end{tablenotes}
      \end{threeparttable}
\end{adjustbox}
\end{table}

\revision{Then, we} run the three variants of \tool (\ie \toolcon, \tooladv and \toolagg) and present the \revision{effects of different driving strategies on the general effectiveness} in Table~\ref{rq4:table7}.

\toolcon ranks first in terms of the proportion of EIVs in the straight road and crossroad scenarios, respectively. In the scenarios generated by \toolcon, NPC vehicles typically travel at lower speeds than the Ego vehicle. However, Apollo exhibits poor performance in predicting slow-moving objects, and its overly conservative following strategy leads to failure in reaching the destination when traveling behind a low-speed NPC vehicle.

\tooladv generates the most EIVs on both straight road and crossroad. This is because the adversarial driving strategy adopted by NPC vehicles effectively increases interactions with the Ego vehicle, raising the difficulty for the Ego vehicle to complete its driving tasks. However, this strategy also leads to more scenarios where the NPC vehicle aggressively collides with a decelerating~Ego vehicle, resulting in a higher false positive rate than~\toolcon.

\toolagg generates the fewest violations. Due to the high speeds of NPC vehicles in its scenarios, most of the NPC vehicles hit the stopped Ego vehicle from behind or from the side aggressively in the reported violations. As a result, the proportion of EIVs found by \toolagg accounts for only \todo{24.10\%} and \todo{16.51\%} on the straight road and crossroad, respectively.

\begin{table}[!t]
    \caption{Effects on the Number of Unique EIV Patterns}
    \vspace{-5pt}
    \label{rq4:table8}
    \centering
    \begin{adjustbox}{width=0.75\linewidth}
\begin{threeparttable}
    \begin{tabular}{cccccc|c}
        \toprule
        \textbf{Method} & \textbf{\#R1$^1$} & \textbf{\#R2$^2$} & \textbf{\#R3$^3$} & \textbf{\#R4$^4$}& \textbf{Sum} & \textbf{Details}\\
        \midrule        
        \textbf{\toolcon}& 7&   3&   3 & 1& \cellcolor{gray!15}14&R1-1, R1-4\textasciitilde R1-9, R2-1\textasciitilde R2-3, R3-1, R3-3, R3-4, R4-1    \\
        \textbf{\tooladv}& 9&   3&  3&1& \cellcolor{gray!15}16& R1-1\textasciitilde R1-7, R1-10, R1-12, R2-1\textasciitilde R2-3, R3-2\textasciitilde R3-4, R4-1    \\
        \textbf{\toolagg}&    4&   1& 1&1&  \cellcolor{gray!15}7&R1-4\textasciitilde R1-7, R2-3, R3-3, R4-1    \\
        \textbf{\tool}& \textbf{12}&   \textbf{3}&   \textbf{4}& \textbf{1} &  \cellcolor{gray!15}\textbf{20} &R1-1\textasciitilde R1-12, R2-1\textasciitilde R2-3, R3-1\textasciitilde R3-3, R4-1\\
        \bottomrule
    \end{tabular}
        \begin{tablenotes}
        \footnotesize
        \item $^1$ \textit{the number of patterns where the Ego vehicle collides with NPC vehicles}.
        \item $^2$ \textit{the number of patterns where the Ego vehicle hits illegal lines}.
        \item $^3$ \textit{the number of patterns where the Ego vehicle gets stuck that fails to reach the destination}.
        \item $^4$ \textit{the number of patterns where the Ego vehicle runs red lights}.
      \end{tablenotes}
\end{threeparttable}
    \end{adjustbox}
\end{table}

In terms of the number of unique EIV patterns, all these three variants can only find part of the patterns discovered by \tool in 12 hours. \tooladv performs best among these three variants. \toolagg finds the fewest unique EIV patterns with only an average of \todo{4.4} ones on the straight road and \todo{1.8} ones on the crossroad. Table~\ref{rq4:table8} shows the total unique patterns of each variant in 5 repetitions of experiments. The result shows that one single driving strategy in scenarios is not conducive to improving the diversity of violation scenarios found.

\begin{tcolorbox}[size=small, opacityfill=1, before skip=5pt, after skip=10pt]
    \textit{\textbf{Summary.}} \revision{The GA-based scenario generator and runtime NPC behavior generation are complementary in \tool. The former improves the effectiveness of scenario initialization, while the latter increases interaction richness during execution. Besides, NPC vehicles employing yielding and adversarial strategies pose greater threats to the Ego vehicle, while the mutation of driving strategies further contributes to the diversity of unique EIV patterns.}
\end{tcolorbox}

\subsection{\revision{RQ5: Sensitivity Analysis of Key Safety Threshold Parameters}}\label{RQ5}

\begin{table}[!t]
    \caption{\revision{Sensitivity of the Key Safety Threshold Parameter}}
    \label{rq5:table9}
    \vspace{-5pt}
    \centering
    \begin{adjustbox}{width=0.6\linewidth}
    \begin{threeparttable}
    {
    \begin{tabular}{cccccc}
        \toprule
        \textbf{Threshold} & \textbf{\#Scenario}$^1$ & \textbf{\#Violation} & \textbf{\#EIV}$^2$ & \textbf{Proportion} & \textbf{\#Unip}$^3$ \\
        \midrule
        \textbf{\tool-20} & 772.2 & 404.8 & 141.9 & 35.05\% & 7.0 \\
        \textbf{\tool-30} & 706.8 & 137.6 & 116.6 & 81.39\% & 9.5 \\
        \textbf{\tool-40} & 512.4 & 82.8 & 78.6 & 94.93\% & 5.5  \\
        \bottomrule
    \end{tabular}
    }
    \begin{tablenotes}
        \footnotesize
        \item $^1$ \textit{the number of the generated scenarios}.
        \item $^2$ \textit{the number of violations induced by the Ego vehicle}.
        \item $^3$ \textit{the number of unique EIV patterns}.
    \end{tablenotes}
    \end{threeparttable}
    \end{adjustbox}
\end{table}

\revision{Table~\ref{rq5:table9} presents the sensitivity analysis results of \tool under different values of the key safety threshold parameter (denoted as \tool-20, \tool-30, and \tool-40, respectively). The reported results are averaged over the straight road and crossroad scenarios, where each scenario type is repeated 5 times.}

\revision{When the threshold is set to 20 meters, \tool finds the largest number of EIVs. Increasing the threshold from 20 meters to 30 meters reduces the average number of EIVs by 17.83\%, and further increasing it to 40 meters leads to a total reduction of 44.61\%. This is because a smaller safety threshold allows NPC vehicles to trigger acceleration, deceleration, and lane-changing maneuvers at a shorter distance from the Ego vehicle, leaving the ADS less reaction time and thus creating more opportunities for violations.}

\revision{However, such close-range interactions also introduce substantially more false positives. Compared with \tool-20, the proportion of EIVs among all reported violations increases by 132.21\% under \tool-30 and by 170.84\% under \tool-40. Equivalently, the false-positive rate decreases from 64.95\% under \tool-20 to 18.61\% under \tool-30 and 5.07\% under \tool-40. This indicates that when the threshold becomes larger, NPC vehicles switch strategies farther away from the Ego vehicle, leaving the ADS more reaction distance and correspondingly reducing overly aggressive or unreasonable interactions. As a result, the reported violations are much more likely to be EIVs.}

\begin{table}[!t]
    \caption{\revision{Unique EIV Patterns under Different Safety Thresholds}}
    \vspace{-5pt}
    \label{rq5:table10}
    \centering
    \begin{adjustbox}{width=0.75\linewidth}
\begin{threeparttable}
    {
    \begin{tabular}{cccccc|c}
        \toprule
        \textbf{Method} & \textbf{\#R1$^1$} & \textbf{\#R2$^2$} & \textbf{\#R3$^3$} & \textbf{\#R4$^4$}& \textbf{Sum} & \textbf{Details}\\
        \midrule        
        \textbf{\tool-20}& 8&   3&   2 & 1& \cellcolor{gray!15}14&R1-2, R1-4\textasciitilde R1-5, R1-8\textasciitilde R1-12, R2-1\textasciitilde R2-3, R3-3\textasciitilde R3-4, R4-1    \\
        \textbf{\tool-40}& 6&   2&  2&1& \cellcolor{gray!15}11& R1-4\textasciitilde R1-5, R1-8\textasciitilde R1-10, R1-12, R2-1, R2-3, R3-3\textasciitilde R3-4, R4-1    \\
        \textbf{\tool-30}& \textbf{12}&   \textbf{3}&   \textbf{4}& \textbf{1} &  \cellcolor{gray!15}\textbf{20} &R1-1\textasciitilde R1-12, R2-1\textasciitilde R2-3, R3-1\textasciitilde R3-3, R4-1\\
        \bottomrule
    \end{tabular}}
        \begin{tablenotes}
        \footnotesize
        \item $^1$ \textit{the number of patterns where the Ego vehicle collides with NPC vehicles}.
        \item $^2$ \textit{the number of patterns where the Ego vehicle hits illegal lines}.
        \item $^3$ \textit{the number of patterns where the Ego vehicle gets stuck that fails to reach the destination}.
        \item $^4$ \textit{the number of patterns where the Ego vehicle runs red lights}.
      \end{tablenotes}
\end{threeparttable}
    \end{adjustbox}
\end{table}

\revision{In terms of the number of unique EIV patterns, \tool-30 performs the best, which is \todo{35.71\%} higher than \tool-20 and 72.73\% higher than \tool-40. This suggests that an excessively small threshold may introduce too many aggressive interactions, many of which repeatedly trigger similar EIVs, while an overly large threshold makes the generated scenarios less challenging and reduces behavioral diversity. Therefore, a moderate threshold can better balance interaction intensity and behavioral reasonableness, thus helping expose more diverse EIV patterns. Table~\ref{rq5:table10} presents the details of the unique EIV patterns found by \tool under different thresholds.}

\revision{Overall, the key safety threshold parameter has a clear impact on the effectiveness of \tool. A smaller threshold helps find more violations, but also introduces substantially more false positives. A larger threshold improves the proportion of EIVs by giving the ADS more reaction distance, but it also reduces the number of discovered violations and weakens the interaction intensity. Among the three settings, 30 meters provides the best trade-off, as it maintains a relatively high EIV proportion, while achieving the highest diversity of unique EIV patterns.}

\begin{tcolorbox}[size=small, opacityfill=1, before skip=5pt, after skip=10pt]
    \revision{\textit{\textbf{Summary.}} The key safety threshold parameter has a clear impact on the effectiveness of \tool. A smaller threshold helps increase interaction density, while a larger threshold improves the EIV proportion but reduces both the number and diversity of discovered EIVs. In our evaluation, the setting of 30 meters provides the best trade-off.}
\end{tcolorbox}

\subsection{\revision{RQ6: Portability Evaluation}}\label{RQ6}

\begin{table}[!t]
    \caption{\revision{Effectiveness and Efficiency of \tool in Roundabout Scenarios using CARLA}}
    \label{rq6:table11}
    \vspace{-5pt}
    \centering
    \begin{adjustbox}{width=0.75\linewidth}
    \begin{threeparttable}
    {
    \begin{tabular}{cccccccc}
        \toprule
         & \textbf{\#Scenario}$^1$ & \textbf{\#Violation} & \textbf{\#EIV}$^2$ & \textbf{Proportion} & \textbf{\#Unip}$^3$ & \textbf{First EIV}$^4$ &  \textbf{One EIV}$^5$ \\
        \midrule
        \textbf{\tool} & 495.2 & 386.8 & 317.4 & 82.06\% & 9.2 & 2.24 & 2.26\\
        \bottomrule
    \end{tabular}
    }
    \begin{tablenotes}
        \footnotesize
        \item $^1$ \textit{the number of the generated scenarios}.
        \item $^2$ \textit{the number of violations induced by the Ego vehicle}.
        \item $^3$ \textit{the number of unique EIV patterns}.
        \item $^4$ \textit{the time used to find the first EIV (minute)}.
        \item $^5$ \textit{the average time used to find one EIV (minute)}.
    \end{tablenotes}
    \end{threeparttable}
    \end{adjustbox}
\end{table}

\revision{Table~\ref{rq6:table11} presents the effectiveness and efficiency of \tool after migrating it to CARLA in roundabout scenarios.}

\revision{The results show that \tool remains effective in the new simulator and more complex traffic environment. Within 12 hours, \tool generates 495.2 scenarios and reports 386.8 violations, among which 317.4 are EIVs, yielding an EIV proportion of 82.06\%. Besides, \tool discovers 9.2 unique EIV patterns on average in this setting. We further analyze the diversity of discovered EIVs in the roundabout scenarios. Although the roundabout we used is different from a standard signalized intersection, its two-lane structure and multiple junction areas still create conflict points similar to those in straight road and intersection scenarios, such as merging, yielding and turning conflicts. Across 5 repetitions, \tool reveals 11 unique EIV patterns, including R1-1, R1-2, R1-3, R1-4, R1-5, R1-9, R1-11, R3-1, R3-2, R3-3, and R3-4.  These results indicate that the core mechanism of \tool, including runtime NPC maneuver decision and online trajectory generation, can still effectively expose ADS weaknesses after being migrated from LGSVL to CARLA.}

\revision{In terms of efficiency, \tool finds the first EIV in 2.24 minutes and requires 2.26 minutes on average to find one EIV. Moreover, the average analysis time and execution time are 0.84 hours and 11.16 hours, respectively. Although the maximum number of NPC vehicles is increased from 4 in LGSVL-based experiments to 8 in CARLA, the additional computational overhead remains limited. This is because \tool performs behavior generation and monitoring asynchronously during simulation execution, which amortizes the computation cost and avoids introducing substantial extra analysis overhead before each run. These results suggest that the dynamic maneuver decision logic and asynchronous execution design of \tool can scale to more complex traffic environments without significantly sacrificing efficiency.}

\revision{Overall, these results demonstrate the portability of \tool across simulators and scenario types. Although our main comparison with prior work is conducted on LGSVL for fairness, \tool can be effectively migrated to CARLA and remains capable of finding EIVs effectively and efficiently in complex scenarios.}

\begin{tcolorbox}[size=small, opacityfill=1, before skip=5pt, after skip=10pt]
    \revision{\textit{\textbf{Summary.}} \tool can be effectively migrated to another simulator and remains both effective and efficient in more complex roundabout scenarios with more NPC vehicles, demonstrating its portability.}
\end{tcolorbox}

\section{Threats to Validity}
First, the selection of target ADS and simulator poses a threat to validity. We select Apollo 8.0 as our target ADS, which is an open-source ADS and widely used in the industry, to ensure fair comparisons with baselines that are also only adapted to this version. \revision{We choose LGSVL rather than CARLA because it has good compatibility with Apollo 8.0 and all the baselines support LGSVL. Newer versions of Apollo are not considered due to unstable bridging with simulators. In addition, we migrate \tool to CARLA to demonstrate its portability.}

Second, the selection of baselines poses another threat to validity. To mitigate this threat, we select \todo{four} state-of-the-art approaches that support Apollo and LGSVL. \textsc{AV-Fuzzer} and BehAVExplor are based on a fuzzing engine using genetic algorithm. \textsc{AutoFuzz} is one of the newest testing approaches guided by neural network and CRISCO is a data-driven approach using influential behavior patterns derived from real-world dataset. It should be noted that the traffic participants in \tool currently support NPC vehicles, while \textsc{AutoFuzz} and CRISCO support pedestrians and cyclists. During the experiments, all approaches only use vehicles for a fair comparison. We do not compare \tool with other approaches due to differences in experimental configurations that can not be fully reproduced, or because they have been compared by our selected baselines. 
DoppelTest~\cite{huai2023doppelganger} aims to find EIVs by 
bridging multiple ADSs together to find violations, which does not use NPCs in the simulator. We do not compare \tool with DoppelTest because it does not work natively with LGSVL and needs to run multiple instances of Apollo concurrently, increasing the computational cost.
All comparisons with baselines show statistically significant differences ($p$ = 0.0079) on the metrics by Mann-Whitney U~test, and the Cohen's d is far greater than 0.8.

Last, the subjective diagnosis and classification of violations induced by the Ego vehicle affects the validity. To mitigate this threat, we use \textsc{DiaVio}~\cite{lu2024diavio}, an LLM-empowered diagnosis approach, and ACAV~\cite{sun2024acav}, a causality-based analysis tool. \revision{After obtaining the preliminary diagnosis results from \textsc{DiaVio}, we asked two additional authors to independently inspect each confirmed EIV and classify it according to the predefined criteria in Sec.~\ref{sec:evaluationsetup}, namely violation type, NPC maneuver category, and underlying causal factor. For collision violations, ACAV was additionally used to support root-cause analysis. When disagreements arose, a third author joined the discussion to reach a consensus. The resulting Cohen's kappa coefficient is 0.862, indicating a high level of inter-rater agreement.}


\section{Related Work}

Scenario-based testing~\cite{zhong2021survey, klischat2022falsifying, li2016intelligence, lou2022testing, Zhang2023, Ding2023, Haq2020} has been widely studied to generate diverse driving scenarios for ADS testing to identify safety violations. A scenario-based safety evaluation framework has been proposed in ISO 34502~\cite{iso34502}. \revision{Some domain specific languages~\cite{Fremont2019, Queiroz2019, openscenario, opendrive, queiroz2024driver} have been proposed to describe the driving scenarios. A few works~\cite{gambi2019generating,bashetty2020deepcrashtest,dai2024sctrans,zhang2023building, guo2024sovar, tang2024legend} attempt to reproduce real-world data (e.g., traffic accident reports and vehicle trajectories) to find corner cases in simulation. To accelerate the identifying of violations, numerous search-based works~\cite{Abdessalem2016a, Abdessalem2016b, li2020av, tian2022mosat, Tian2022} use a genetic algorithm-based approach to generate scenarios where the Ego vehicle may collide with NPC vehicles, while a few works~\cite{abdessalem2018testinga, Gladisch2020, luo2021targeting, kim2022drivefuzz, cheng2023behavexplor, Huai2023a} guide the ADSs to violate predefined rules, such as failing to reach their destinations, or to exhibit incorrect behaviors like speeding or executing unsafe lane changes. Some works~\cite{sun2022lawbreaker, zhang2023testing, li2023simulation,li2024viohawk} propose to generate driving scenarios where ADSs break specific traffic rules. Additionally, Lu et al.~\cite{lu2023test}, Zhong et al.~\cite{zhong2022neural} and Wang et al.~\cite{wang2024dance} employ neural network to guide the generation of scenarios. Besides, several works~\cite{birchler2024does, woodlief2024s3c, neelofar2024towards} investigate the metrics (e.g., physical environment-state coverage metric~\cite{hildebrandt2023physcov}) in simulation to guide the search and Chen et al.~\cite{chen2024misconfiguration} study the configuration of simulation in ADS testing.}

\revision{However, many prior scenario-based testing approaches, especially search-based ones, determine most scenario elements before execution, including NPC behaviors. They may insufficiently capture interactions conditioned on the Ego vehicle's real-time behavior and traffic signals during simulation, resulting in NPC vehicles not obeying traffic rules (\eg traffic lights) and performing unreasonable behaviors. In contrast, \tool dynamically generates NPC maneuvers and trajectories during execution, which helps find more violations induced by the ADS. Huai et al.~\cite{huai2023doppelganger} also aim to maximize violations induced by the ADS. They bridge multiple ADSs for interaction using Apollo's own simulation module \textit{SimControl}, rather than using NPC vehicles in the simulator. However, this is achieved at the cost of feeding ground truth directly into the localization and perception modules of Apollo. Differently, we propose a new framework to generate scenarios in the simulator and test ADSs at the system-level.}

\revision{Several recent works further explore reinforcement learning to generate interactive and adversarial NPC behaviors during execution~\cite{rowe2024ctrl, rempe2022generating, doreste2024adversarial}. Unlike traditional search-based approaches that manipulate static scenario elements, these works model surrounding vehicles as independent adversarial agents and train them via reinforcement learning to actively challenge the Ego vehicle during simulation, maximizing the likelihood of exposing violation scenarios. For example, Wang et al.~\cite{wang2025amacollision} propose an adversarial multi-agent framework that generates attack-oriented traffic behaviors to efficiently uncover safety violations. However, these approaches usually rely on learned policies and carefully designed rewards, which require substantial training data and computational resources. Differently, \tool adopts a rule-based mechanism that does not require policy training, providing controllable and interpretable NPC behaviors. Moreover, \tool tightly integrates runtime NPC behavior generation with search-based scenario generation and execution over driving tasks, weather conditions, and traffic signal configurations, enabling effective and efficient system-level testing for finding more violations~induced~by~the~ADS.}


\section{Conclusion}
We have proposed and implemented \tool, a novel \revision{search-based testing framework}, to find more violation scenarios induced by the ADS through dynamically generating NPC vehicle behaviors during simulation execution. Large-scale experiments have been conducted to demonstrate the effectiveness and efficiency of \tool. In future, we plan to extend \tool to support more ADSs and simulators. Moreover, we also plan to support more types of traffic participants (\ie pedestrians and cyclists). The experimental records and source code of our work is available at our replication site~\cite{website}.

\begin{acks}
This work was supported by the National Natural Science Foundation of China (Grant No. 92582205).
\end{acks}

\bibliographystyle{ACM-Reference-Format}
\bibliography{src/reference}

\end{document}